\documentclass[conference]{IEEEtran}

\usepackage{makecell}
\usepackage{subcaption}
\usepackage{longtable}
\usepackage[utf8]{inputenc}
\usepackage{amsmath} 
\usepackage{lipsum}
\usepackage{float}
\usepackage{graphicx}
\usepackage{subcaption} 
\usepackage{pdflscape}
\usepackage{placeins}
\usepackage{listings}
\usepackage{xcolor}
\usepackage{adjustbox}  
\usepackage{array}  
\usepackage{multirow}
\usepackage{todonotes}
\usepackage{siunitx}
\usepackage{listings}
\usepackage{caption}
\usepackage{threeparttable}
\usepackage{colortbl}
\usepackage{balance}
\usepackage{pifont}

\newcommand{\activeVulnsNum}{\num{24}}
\newcommand{\papersNum}{\num{71}}
\newcommand{\totalLosses}{\num{1093979779}}

\usepackage[font=small,skip=2pt]{caption}

\definecolor{ourwork}{RGB}{33, 150, 243}    
\definecolor{gray1}{RGB}{245, 245, 245}

\usepackage{rotating}
\usepackage{array}
\usepackage{longtable}
\usepackage{tabularx}
\usepackage{amssymb}
\usepackage{booktabs}
\usepackage{enumitem}
\usepackage{etoolbox}
\usepackage{xurl}

\usepackage{inconsolata}
\DeclareRobustCommand{\code}[1]{\lstinline[basicstyle=\ttfamily\small]{#1}}

\usepackage{hyperref}
\usepackage[nameinlink,noabbrev]{cleveref}
\crefname{insightcounter}{Insight}{Insights}   
\Crefname{insightcounter}{Insight}{Insights}   
\crefname{insight}{Insight}{Insights}

\definecolor{verylightgray}{rgb}{.97,.97,.97}

\lstdefinelanguage{Solidity}{
    showstringspaces=false, 
	keywords=[1]{anonymous, assembly, assert, balance, break, call, callcode, case, catch, class, constant, continue, constructor, contract, debugger, default, delegatecall, delete, do, else, emit, event, experimental, export, external, false, finally, for, function, gas, if, implements, import, in, indexed, instanceof, interface, internal, is, length, library, log0, log1, log2, log3, log4, memory, modifier, new, payable, pragma, private, protected, public, pure, push, require, return, returns, revert, selfdestruct, send, solidity, storage, struct, suicide, super, switch, then, this, throw, transfer, true, try, typeof, using, value, view, while, with, addmod, ecrecover, keccak256, mulmod, ripemd160, sha256, sha3},
	keywordstyle=[1]\color{blue}\bfseries,
	keywords=[2]{address, bool, byte, bytes, bytes1, bytes2, bytes3, bytes4, bytes5, bytes6, bytes7, bytes8, bytes9, bytes10, bytes11, bytes12, bytes13, bytes14, bytes15, bytes16, bytes17, bytes18, bytes19, bytes20, bytes21, bytes22, bytes23, bytes24, bytes25, bytes26, bytes27, bytes28, bytes29, bytes30, bytes31, bytes32, enum, int, int8, int16, int24, int32, int40, int48, int56, int64, int72, int80, int88, int96, int104, int112, int120, int128, int136, int144, int152, int160, int168, int176, int184, int192, int200, int208, int216, int224, int232, int240, int248, int256, mapping, string, uint, uint8, uint16, uint24, uint32, uint40, uint48, uint56, uint64, uint72, uint80, uint88, uint96, uint104, uint112, uint120, uint128, uint136, uint144, uint152, uint160, uint168, uint176, uint184, uint192, uint200, uint208, uint216, uint224, uint232, uint240, uint248, uint256, var, void, ether, finney, szabo, wei, days, hours, minutes, seconds, weeks, years},
	keywordstyle=[2]\color{teal}\bfseries,
	keywords=[3]{block, blockhash, coinbase, difficulty, gaslimit, number, timestamp, msg, data, gas, sender, sig, value, now, tx, gasprice, origin},
	keywordstyle=[3]\color{violet}\bfseries,
	identifierstyle=\color{black},
	sensitive=false,
	comment=[l]{//},
	morecomment=[s]{/*}{*/},
	commentstyle=\color{gray}\ttfamily,
	stringstyle=\color{red}\ttfamily,
	morestring=[b]',
	morestring=[b]",
}

\makeatletter
\pretocmd{\lst@inline}{\let\lst@basicstyle\ttfamily\normalsize}{}{}
\makeatother

\AtBeginDocument{%
  }

\lstdefinestyle{solidity-common}{
  language       = Solidity,
  keywordstyle   = \color{blue}\bfseries,
  ndkeywordstyle = \color{teal}\bfseries,
  commentstyle   = \color{gray}\itshape,
  stringstyle    = \color{red},
  backgroundcolor= \color{gray!10},
  tabsize        = 2,
  breaklines     = true,
  breakatwhitespace = true,
  captionpos     = b
}

\usepackage[most]{tcolorbox}
\newtcolorbox[auto counter]{rqbox}[2][]{
  colback=blue!3!white, colframe=blue!25!black,
  coltitle=black, fonttitle=\bfseries,
  colbacktitle=blue!10!white, 
  title={\textbf{#2}}, 
  sharp corners, 
  boxrule=0.3mm, 
  width=\linewidth, 
  enhanced,
  drop shadow,
  left=1mm, right=1mm, top=1mm, bottom=1mm,
  #1
}

\newcounter{insight}
\newtcolorbox{ActualResultBox}[2][]{%
  colback=teal!3!white,
  colframe=teal!25!black,
  coltitle=black,
  fonttitle=\bfseries,
  colbacktitle=teal!10!white,
  title={#2},
  before upper={\ifx&#1&\else\label{#1}\fi},
  sharp corners,
  boxrule=0.3mm,
  enhanced,
  left=1mm, right=1mm, top=1mm, bottom=1mm,
  before skip=0pt, 
  after skip=0pt,  
  boxsep=0pt,      
  breakable,       
  floatplacement=!ht,
}
\newtcolorbox{resultbox}[2][]{%
  colback=teal!3!white,
  colframe=teal!25!black,
  coltitle=black,
  fonttitle=\bfseries,
  colbacktitle=teal!10!white,
  before upper={\ifx&#1&\else\label{#1}\fi},
  title={\refstepcounter{insight}\textbf{Insight \theinsight: #2}},
  sharp corners,
  boxrule=0.3mm,
  width=\linewidth,
  enhanced,
  drop shadow,
  left=1mm,right=1mm,top=1mm,bottom=1mm
}
\newcommand{\insightref}[1]{Insight~\ref{#1}}

\newcounter{insightcounter}
\newtcolorbox{InsightBox}[2][]{%
  colback=teal!3!white,
  colframe=teal!25!black,
  coltitle=black,
  fonttitle=\bfseries,
  colbacktitle=teal!10!white,
  title={\refstepcounter{insightcounter}%
         \textbf{Insight \theinsightcounter: #2}%
         \ifx\\#1\\\else\label{#1}\fi},
  sharp corners,
  boxrule=0.2mm,
  width=\linewidth,
  enhanced,
  drop shadow,
  left=0.3mm,right=0.3mm,top=0.3mm,bottom=0.3mm,
}

\begin{document}



\title{SoK: Root Causes of \$1 Billion Loss in Smart Contract Real-World Attacks via a Systematic Literature Review of Vulnerabilities}

\author{
\IEEEauthorblockN{Hadis Rezaei}
\IEEEauthorblockA{University of Salerno\\
hrezaei@unisa.it}
\and
\IEEEauthorblockN{Mojtaba Eshghie}
\IEEEauthorblockA{KTH Royal Institute of Technology\\
eshghie@kth.se}
\and
\IEEEauthorblockN{Karl Andersson}
\IEEEauthorblockA{Luleå University of Technology\\
karl.andersson@ltu.se}
\and
\IEEEauthorblockN{Francesco Palmieri\\}
\IEEEauthorblockA{University of Salerno\\
fpalmieri@unisa.it}
}

\author{
\IEEEauthorblockN{Hadis Rezaei\textsuperscript{*}}
\IEEEauthorblockA{University of Salerno\\
hrezaei@unisa.it}
\and
\IEEEauthorblockN{Mojtaba Eshghie\textsuperscript{*}}
\IEEEauthorblockA{KTH Royal Institute of Technology\\
eshghie@kth.se}
\and
\IEEEauthorblockN{Karl Andersson}
\IEEEauthorblockA{Luleå University of Technology\\
karl.andersson@ltu.se}
\and
\IEEEauthorblockN{Francesco Palmieri\\}
\IEEEauthorblockA{University of Salerno\\
fpalmieri@unisa.it}
\thanks{*Joint first authors with equal contribution.}
}


\IEEEoverridecommandlockouts
\makeatletter\def\@IEEEpubidpullup{6.5\baselineskip}\makeatother
\IEEEpubid{\parbox{\columnwidth}{
		Network and Distributed System Security (NDSS) Symposium 2025\\
		24-28 February 2025, San Diego, CA, USA\\
		ISBN 979-8-9894372-8-3\\
		https://dx.doi.org/10.14722/ndss.2025.[23$|$24]xxxx\\
		www.ndss-symposium.org
}
\hspace{\columnsep}\makebox[\columnwidth]{}}

\makeatletter
\renewcommand{\paragraph}[1]{%
  \vspace{0.2ex}%
  \noindent\textbf{#1.}%
  \hspace{0.5em}%
  \@afterheading%
  \vspace{0.2ex}%
}
\makeatother

\maketitle

\begin{abstract}
While catastrophic attacks on Ethereum persist, vulnerability research remains fixated on implementation-level smart contract bugs, creating a gap between academic understanding of vulnerabilities and the true root causes of high-impact, real-world incidents.
To address this, we employ a two-pronged methodology: first, a systematic literature review of {71} academic papers to build a catalog of {24} active and {5} deprecated vulnerabilities. Second, we conduct an in-depth, empirical analysis of {50} of the most severe real-world attacks between 2022 and 2025, collectively incurring over \$1.09B in losses, to identify their root causes. We introduce the concept of ``exploit chains'' by revealing that many incidents are not caused by isolated vulnerabilities but by combinations of human, operational, and economic design flaws that link with implementation bugs to enable an attack.
Our analysis yields insights on \emph{how decentralized applications are exploited in practice}, leading to a novel, four-tier root-cause framework that moves beyond code-level vulnerabilities. We find that real-world successful attacks on Ethereum (and related networks) trace back to one of the four tiers of (1) protocol logic design, (2) lifecycle and governance, (3) external dependencies, and (4) classic smart contract vulnerabilities. We investigate the suitability of this multi-tier incident root-cause framework via a case study. 
\end{abstract}


%
\IEEEpeerreviewmaketitle

\section{Introduction}

The Ethereum blockchain and its smart contract ecosystem have grown to manage hundreds of billions of dollars~\cite{EthMarketCap,ethereum2014}. Along with this growth, in recent years, a relentless series of successful high-profile hacks has occurred, resulting in staggering financial losses~\cite{zhou2023sok,OracleExploitation}. Some reports suggest up to \$\num{2.2} billion stolen from protocols in 2024~\cite{chainalysis}.
In response to such attacks, a substantial body of research has been performed to classify and detect smart contract vulnerabilities. These efforts focused on implementation-level bugs such as reentrancy and integer overflows~\cite{luu2016making, atzei2017survey}. This led to the development of numerous static and dynamic analysis tools like Slither~\cite{feist2019slither} and Echidna~\cite{Echidna} or even execution frameworks to run a swarm of analysis tools on a contract, such as SmartBugs~\cite{SmartBugs2}. Despite their usefulness in detecting certain classes of vulnerabilities~\cite{EmpricalReview}, both the severity (total funds stolen) and number of successful attacks have increased in recent years~\cite{chainalysis}.

\begin{table}[!t]
\caption{Comparison of smart contract security SLRs and SoKs.}
\label{tab:survey_matrix}
\centering
\tiny
\setlength{\tabcolsep}{1pt}
\renewcommand{\arraystretch}{0.9}
\begin{tabular}{@{}p{0.8cm}p{0.4cm}ccccp{4.5cm}@{}}
\toprule
\textbf{Paper} & \textbf{Year} & \textbf{Vulns} & \textbf{Prev.}$^b$ & \textbf{Tax.}$^a$ & \textbf{Wild}$^c$ & \textbf{Key Strength} \\
\midrule
\rowcolor{gray1}
\makebox[0pt][l]{\hypertarget{row:S1}{}}S1~\cite{chen2020survey} & 2020 & 40 & P & 4L & \textcolor{red!70!black}{\ding{55}} & Vulnerability $\leftrightarrow$ attack mapping \\
\makebox[0pt][l]{\hypertarget{row:S2}{}}S2~\cite{de2021vulnerabilities} & 2021 & 20 & \textcolor{red!70!black}{\ding{55}} & 3C & \textcolor{red!70!black}{\ding{55}} & Comprehensive tool analysis \\
\rowcolor{gray1}
\makebox[0pt][l]{\hypertarget{row:S3}{}}S3~\cite{samreen2021survey} & 2021 & 8 & \textcolor{red!70!black}{\ding{55}} & NI & C & Maps vulnerabilities $\rightarrow$ attacks $\rightarrow$ detection tools \\
\makebox[0pt][l]{\hypertarget{row:S4}{}}S4~\cite{tolmach2021survey} & 2021 & 202 & \textcolor{green!70!black}{\ding{51}} & FM & \textcolor{red!70!black}{\ding{55}} & Comprehensive formal methods for all SC$^i$ verification. \\
\rowcolor{gray1}
\makebox[0pt][l]{\hypertarget{row:S5}{}}S5~\cite{kissoon2022detecting} & 2022 & 6 & \textcolor{red!70!black}{\ding{55}} & DT & \textcolor{red!70!black}{\ding{55}} & Vulnerability detection tools quality gaps \\
\makebox[0pt][l]{\hypertarget{row:S6}{}}S6~\cite{kushwaha2022systematic} & 2022 & 24 & \textcolor{red!70!black}{\ding{55}} & RS & \textcolor{red!70!black}{\ding{55}} & Causal hierarchy enabling systematic prevention \\
\rowcolor{gray1}
\makebox[0pt][l]{\hypertarget{row:S7}{}}S7~\cite{chu2023survey} & 2023 & 12 & \textcolor{red!70!black}{\ding{55}} & 3L & \textcolor{red!70!black}{\ding{55}} & Layer-based classification: Solidity, EVM, Blockchain \\
\makebox[0pt][l]{\hypertarget{row:S8}{}}S8~\cite{porkodi2024smart} & 2023 & 6 & \textcolor{red!70!black}{\ding{55}} & TB & \textcolor{red!70!black}{\ding{55}} & Categorizes 6 vulns (BE/RC/EX/ED/BD)$^g$ + tool assessment \\
\rowcolor{gray1}
\makebox[0pt][l]{\hypertarget{row:S9}{}}S9$^*$~\cite{zhou2023sok} & 2023 & 50+ & \textcolor{green!70!black}{\ding{51}} & 5L & 181 & Analyzes 181 exploits across NET/CON/SC/PRO/AUX$^e$ layers \\
\makebox[0pt][l]{\hypertarget{row:S10}{}}S10~\cite{vidal2024openscv} & 2024 & 94 & \textcolor{red!70!black}{\ding{55}} & HR & E & Open hierarchical taxonomy with code examples \& ODC$^h$ \\
\rowcolor{gray1}
\makebox[0pt][l]{\hypertarget{row:S11}{}}S11$^*$~\cite{ruggiero2024sok} & 2024 & V$^d$ & \textcolor{red!70!black}{\ding{55}} & UN & \textcolor{red!70!black}{\ding{55}} & ER$^j$ data model for unifying SC$^i$ vulnerability taxonomies \\
\makebox[0pt][l]{\hypertarget{row:S12}{}}S12$^*$~\cite{augusto2024sok} & 2024 & 45 & \textcolor{green!70!black}{\ding{51}} & 4L & 92 & Traces vulnerabilities in real attacks \\
\rowcolor{gray1}
\makebox[0pt][l]{\hypertarget{row:S13}{}}S13~\cite{islam2025securing} & 2025 & 20+ & \textcolor{green!70!black}{\ding{51}} & 5L & \textcolor{red!70!black}{\ding{55}} & Comprehensive 5-layer attack classification \\
\midrule
\rowcolor{ourwork}
\textcolor{white}{\textbf{Our Work}} & \textcolor{white}{\textbf{2025}} & \textcolor{white}{\textbf{24+5}$^f$} & \textcolor{white}{\textbf{\textcolor{green!70!black}{\ding{51}}}} & \textcolor{white}{\textbf{Cause}} & \textcolor{white}{\textbf{50}} & \textcolor{white}{\textbf{Real-world validation, root-cause framework from incidents}} \\
\bottomrule
\end{tabular}
\vspace{0.1cm}
{\tiny
\noindent\textbf{Footnotes:} 
$^*$SoK
$^a$\textbf{Tax:} xL=x-Layer, xC=x-Category FM=Formal Methods, DT=Detection-based, RS=Root-cause-based, TB=Tool-Based, HR=Hierarchical, UN=Unified • 
$^b$\textbf{Prev:} \textcolor{green!70!black}{\ding{51}}=comprehensive, P=partial, \textcolor{red!70!black}{\ding{55}}=none • 
$^c$\textbf{Wild:} Numbers=exploit count, C=case studies, E=examples, \textcolor{red!70!black}{\ding{55}}=none • 
V$^d$=Various/unspecified count • $^e$NET/CON/SC/PRO/AUX=Network/Consensus/Smart Contract/Protocol/Auxiliary • $^f$\activeVulnsNum\ core + 5 deprecated vulnerabilities • 
$^g$BE=Bugs/Errors, RC=Rogue Contracts, EX=Extortion, ED=External Data, BD=Backdoors • 
$^h$ODC=Orthogonal Defect Classification • 
$^i$SC=Smart Contract • 
$^j$ER=Entity-Relationship
}
\end{table}

\paragraph{\textbf{Related SoKs and Surveys}}\Cref{tab:survey_matrix} presents a review of three Systematization of Knowledge (SoK) and ten Systematic Literature Review (SLR) papers on Decentralized Application (DApp) security. 
Surveys and SoKs classify tens of vulnerabilities across different layers of the blockchain stack (e.g., network, smart contract, and protocol layers)~\cite{chen2020survey,kushwaha2022systematic, chu2023survey}. These taxonomies have become increasingly granular in classifying weaknesses~\cite{kushwaha2022systematic} and mapping them to formal verification properties~\cite{tolmach2021survey}, or aligning them with frameworks such as NIST bugs~\cite{samreen2021survey}. However, our review unveils a few gaps in this literature: First, \emph{prevention gap}: only 4 of 13 works (\hyperlink{row:S4}{S4}, \hyperlink{row:S9}{S9}, \hyperlink{row:S12}{S12}, and \hyperlink{row:S13}{S13}) address prevention methods, most of which focus exclusively on detection. Second, the \emph{validation gap}: merely 3 works validate their findings against real-world exploits—\hyperlink{row:S3}{S3 (case studies)}, \hyperlink{row:S9}{S9 (181 exploits)} and \hyperlink{row:S12}{S12 (92 exploits)}, while others \emph{rely on synthetic datasets}. Third, the \emph{taxonomy evolution} shows a shift from simple categorization to sophisticated multi-layer frameworks (4L, 5L), \emph{yet none adopt a root-cause approach based on real incidents}.

\paragraph{\textbf{Telltale Signs from Real-World Incidents}}
At 08:56 UTC on 13 March 2023, the Euler Finance lending market was solvent; four minutes later, it was \$\num{197} million in deficit~\cite{rekt2023euler,slowmist2023euler}. Nothing in its Solidity source code could be attributed to common smart contract vulnerabilities, no unchecked casts, no arithmetic overflow, or no reentrancy bug. Instead, an unguarded auxiliary function, a liquidation discount, and a flash loan combined into an unexpected profit engine to exploit the contract. Several months earlier, the Nomad bridge was emptied of \$\num{190}M because a single initializer function had been called with the wrong argument during an upgrade, resetting its trusted root to zero address and turning every message into valid~\cite{nomad_rekt_2022}. 
Episodes like these are not rare accidents~\cite{chainalysis}. The vast majority of existing literature and automated tooling remains heavily focused on implementation-level weaknesses, the ``known unknowns''~\cite{chen2020survey,kushwaha2022systematic,tolmach2021survey}. Yet, as incidents like the \$\num{197}M Euler Finance exploit~\cite{rekt2023euler} show, financially devastating attacks may not merely stem from coding mistakes. In the mentioned protocol's case, the attack succeeded as the liquidation mechanism was flawed. The code executed exactly as intended; it was the intention itself that created the vulnerability.
This incident is not an isolated case. It highlights a systemic issue: while literature focuses on finding bugs in the code, a large-scale analysis of real-world attacks and comparing the exploited vulnerabilities in these incidents and their alignment with the literature is required. Thus, we propose the following research questions:

\begin{enumerate}[nosep,leftmargin=*]
\item \textbf{RQ1:} Which vulnerabilities \emph{still} affect Ethereum smart contracts as of \num{2025}, and how can they be mitigated?
\item \textbf{RQ2:} What factors, other than mere coding vulnerability patterns (identified in RQ1), are evident in the real-world incidents and contribute to the success of the attacks?
\item \textbf{RQ3:} What categories do these vulnerabilities fall into based on their root causes?
\end{enumerate}

To address RQ1, we conduct a systematic literature review (SLR) of \papersNum\ seminal papers, synthesizing their findings into a comprehensive catalog of \activeVulnsNum\ active and \num{5} deprecated vulnerabilities. This provides a robust, state-of-the-art baseline of academically recognized threats. To address RQ2, we perform a deep, qualitative analysis of 50 high-impact smart contract exploits that occurred between 2022 and early 2025, representing a total of \$\totalLosses\ in losses. By examining post-mortem reports, on-chain transaction data, and the attacked protocols' source code and documentations, we move beyond bug labels (e.g., reentrancy) to uncover the steps (\emph{exploit chains}~\Cref{sec:rq2-results}) that enabled these attacks.
This approach yielded a novel, \emph{four}-tier \emph{root-cause} framework derived directly from our empirical analysis, complementing the academic literature (\Cref{sec:rq3-results}). We argue that real-world exploits are best understood as originating from one these four tiers: 1) flawed economic design and protocol logic (\num{24}\%), 2) protocol lifecycle and governance failures (\num{24}\%), 3) external dependency vulnerabilities (\num{24}\%), and 4) implementation-level weaknesses (\num{28}\%).

In summary, this paper makes the following contributions:
\begin{enumerate}[nosep,leftmargin=*]
\item An up-to-date, curated catalog of \activeVulnsNum\ \emph{active} and \num{5} \emph{deprecated} smart contract vulnerabilities based on a systematic review of \papersNum\ papers. For each active vulnerability type we include detailed prevention/mitigation strategies and real-world incidents involving (in the accompanying repository\footnote{\href{https://github.com/HadisRe/SoK-Root-Cause-of-Smart-Contract-Incidents}{https://github.com/HadisRe/SoK-Root-Cause-of-Smart-Contract-Incidents}}). 
\item A manual analysis of \num{50} high-profile successful incidents\footnote{Distinct incidents from the our real-world examples in contribution 1.} (selected from a pool of \num{337} incidents) occurred in 2022--2025 on various protocol types (e.g., DeFi, cross-chain bridges, etc.) demonstrating that the most {severe real-world exploits stem from factors beyond mere/isolated implementation mistakes}. We identify \emph{exploit chains} that combine human, operational, and economic design flaws with implementation bugs to enable attack vectors. 
\item A novel, \emph{four-tier root-cause} framework for smart contract exploits, grounded on our analysis of real-world exploit methods.
\item A case study on a sub-category of our root-cause-based framework with defense strategies against exploits targeting this sub-category.
\item A dataset of real-world successful attack attributes, including exploited vulnerability types (confirmed by manual expert analysis), victim and target account addresses, and attack transaction hashes.
\end{enumerate}

By systematizing the knowledge of \textcolor{blue!50!black}{\emph{how smart contracts fail in \textbf{practice}}}{{ (and attributing these failures to vulnerabilities found in our SLR and root-causes in our four-tier framework)}}, we show that many successful attacks stem not just from code-level bugs, but from a combination of design, governance, dependency, and operational factors. The rest of the paper is organized as follows. \Cref{sec:methodology} presents our rigorous SLR and incident review protocol. \Cref{sec:rq1-results} presents the results of our SLR for RQ1. \Cref{sec:rq2-results} presents the results of our real-world incident analysis insights for RQ2. \Cref{sec:rq3-results} demonstrates our multi-tier incident root-cause framework for RQ3. Finally, \Cref{sec:conclusion} concludes the paper.

\small{ }

\section{Methodology}\label{sec:methodology}

To answer RQ1–RQ3, we analyzed both academic literature and high-quality technical smart contract incident (successful attack) reports. \Cref{sec:lit-study-protocol,sec:incident-review-protocol} elaborate on the protocol used to identify, collect, and review these resources.

\subsection{Academic Literature Study Protocol}\label{sec:lit-study-protocol}

\begin{figure*}[!t]
    \centering
    \includegraphics[width=0.85\textwidth]{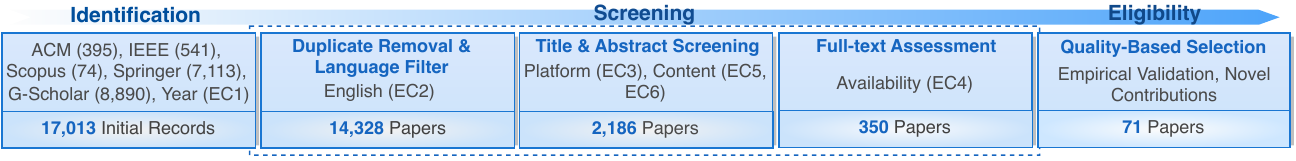}
    \footnotesize
     \caption{\footnotesize
     Systematic review filtering process showing the reduction from \num{17013} identified papers to \papersNum\ quality-validated papers.}
       \label{fig:searchResults}
\end{figure*}

To answer RQ1, we conducted a systematic literature review (SLR) to obtain a comprehensive and structured understanding of Ethereum smart contract vulnerabilities from academic literature. Following Kitchenham`’s guidelines~\cite{kitchenham2007guidelines}, our review process includes study identification (\Cref{sec:study-identification}) and study selection (\Cref{sec:study-selection}).

\subsubsection{\textbf{Study Identification}}\label{sec:study-identification}

Here, we first present the process of formulating search queries for scientific databases. Then, we present the process of filtering them to collect relevant scientific literature from reputable sources (briefly in \Cref{fig:searchResults}). 

\paragraph{Search Keywords} We used the PICOC (Population, Intervention, Comparison, Outcome, Context) framework~\cite{kitchenham2007guidelines} to achieve a systematic search. According to this method, we defined keywords for our search in \Cref{tab:keywordsExtraction}. The ``Population'' was defined as the technology (smart contracts, blockchain), and the ``Intervention'' as the problem being investigated.

\begin{table}[!h]
\scriptsize
\setlength{\tabcolsep}{3pt}
\centering
\caption{Keyword extraction method}
\label{tab:keywordsExtraction}
\begin{tabular}{p{0.3cm} p{2.8cm} p{5cm}}
\toprule
\textbf{Cat.} & \textbf{Keyword} & \textbf{Synonyms} \\
\midrule
P & ``Ethereum smart contracts'' & ``Ethereum contracts'',  ``Solidity'' \\
I & ``Vulnerabilities'' & ``Security Flaws'' OR ``Security Weaknesses'' \\
O & ``Mitigation'' & ``Classification'', ``Mitigation'', ``Prevention'', ``Categorization'', ``classification'', ``taxonomy'' \\
\bottomrule
\end{tabular}
\end{table}

For search keywords, we added synonyms to the main keywords for varied searches. The final query for searching in scientific databases is the result of concatenating all keywords for \emph{P}, \emph{I}, and \emph{O} (and their synonyms) from \Cref{tab:keywordsExtraction}. 

\paragraph{Study Sources} We queried five databases: ACM, IEEE Xplore, Scopus, Springer, and Google Scholar (for pre-prints, theses, and gray literature) to collect studies. \Cref{fig:searchResults} presents the number of papers retrieved from each: ACM (\num{395}), IEEE (\num{541}), Scopus (\num{74}), Springer (\num{7113}), and Google Scholar (\num{8890}). 

\subsubsection{\textbf{Study Selection}}\label{sec:study-selection}

Following the initial identification of \num{17013} papers, we removed duplicates and filtered them based on the exclusion criteria in \Cref{tab:exclusionCriteria}. Specifically, language (EC2) and publication year 2018--2024 (EC1) filtering reduced the count to \num{14328} papers. Title and abstract screening using criteria EC3 (Ethereum focus), EC5, and EC6 (topic relevance) yielded \num{2186} papers. Full-text assessment based on availability (EC4) further refined the selection to \num{350} papers. Finally, quality-based evaluation via manual review of the paper and identification of novel contributions resulted in \papersNum\ papers for in-depth analysis.
\Cref{fig:searchResults} illustrates this process.

\subsection{Real-World Incident Review Protocol}\label{sec:incident-review-protocol}
\begin{figure}[!ht]
\centering\includegraphics[width=1\columnwidth]{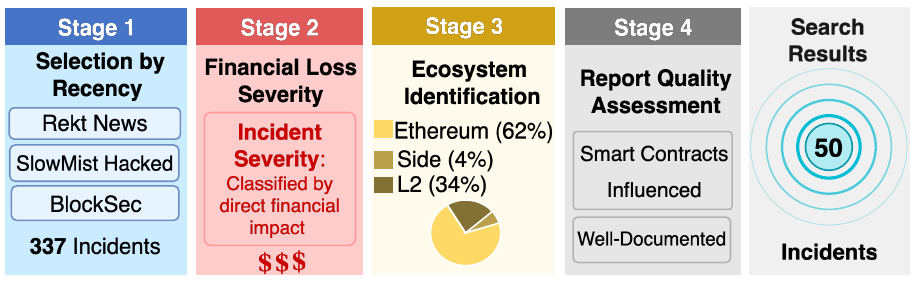}
\caption{
\footnotesize
Four-stage smart contract incident review protocol to systematically identify, prioritize, and analyze relevant attacks.}
\label{fig:incident_protocol}
\end{figure}

To answer RQ2, we reviewed post-mortem incident reports. To ensure the relevance of the successful hacks and maintain quality control over the reviewed resources, we applied a four-stage incident review protocol in \Cref{fig:incident_protocol}.

\noindent
\textbf{Stage 1: Comprehensive Data Collection.} We aggregated incident reports from four reliable blockchain security sources: Rekt News~\cite{rekt_news}, REKT database~\cite{defiyield_rekt}, SlowMist Hacked database~\cite{slowmist_hacked}, and BlockSec reports~\cite{blocksec_blog}. From each source, we extracted up to 100 of the most recent attacks based on the occurrence date. The report sources are selected for their report quality and {analytical depth}. In BlockSec's case, our temporal parameters (April 2022 to April 2025) yielded only 37 documented attacks meeting our criteria, all of which were incorporated into our analysis. As a result, our initial dataset comprised a total of \num{337} attacks for analysis.

\noindent{\textbf{Stage 2: Impact-Based Prioritization.}} Incidents underwent prioritization based on financial impact severity, emphasizing attacks resulting in substantial monetary losses. This criterion ensured our analysis concentrated on vulnerabilities manifesting the most significant ecosystem consequences.

\noindent{\textbf{Stage 3: Ecosystem Classification.}} 
We focused on incidents occurring within the Ethereum blockchain and its compatible networks. Each case was evaluated for relevance, resulting in a final dataset of \num{50} distinct incidents that offer sufficient breadth while maintaining ecosystem coherence. We then mapped each incident to its respective blockchain environment, as shown in the network column of \Cref{tab:main_attack_attributes}, establishing three network categories: 

\begin{enumerate}[leftmargin=*, itemsep=0pt, parsep=0pt, topsep=0pt, partopsep=0pt,labelindent=0pt]
   \item \textbf{Ethereum Mainnet:} The Ethereum network where native consensus exist, identified by transactions on \href{https://etherscan.io}{etherscan.io} (\num{31} attacks).
   \item \textbf{Ethereum Layer 2 Solutions: }These include Arbitrum, Optimism, Base, zkSync, Linea, Scroll, StarkNet, and Mode (\num{17} attacks). L2s inherit security from Ethereum while processing transactions off-chain~\cite{base2025bridges}. They use either Optimistic or Zero-Knowledge rollups to post transaction data back to Layer 1.
   \item \textbf{Ethereum Side-Chains}: Networks like Polygon, BSC (Binance Smart Chain), and Ronin that maintain independent consensus mechanisms while supporting Ethereum Virtual Machine (EVM) compatibility~\cite{ethereum2025scaling} (\num{2} attacks).
\end{enumerate}

This classification follows the technical distinction that true L2 solutions derive their security from Ethereum itself, while side-chains operate with separate validator sets and consensus mechanisms. For example, Base is considered an L2 because it uses the OP Stack to inherit Ethereum's security properties, whereas Polygon PoS uses its own validator set for consensus, making it a side-chain.%

\begin{table}[!t]
\centering
\begin{threeparttable}
\caption{List of exclusion criteria}\label{tab:exclusionCriteria}
\begin{tabular}{ @{} p{0.25cm} p{8cm} }
\hline
\textbf{ID} & \textbf{Exclusion Criteria} \\
\hline
EC1 & The article was published before 2018.\tnote{a} \\
EC2 & The article is written in a language other than English. \\
EC3 & The article focuses on vulnerabilities in platforms other than Ethereum. \\
EC4 & There is no full-text version of the article available. \\
EC5 & The article is a book chapter or editorial notes. \\
EC6 & The article does not concern any of these topics: survey of smart contract vulnerabilities, security of smart contracts, vulnerability detection or mitigation methods for smart contracts, demonstrates findings for smart contracts through simulation or empirical evidence.\\ 
\hline
\end{tabular}
\begin{tablenotes}
\footnotesize
\item[a] The vulnerabilities before \num{2018} are included via more recent papers.
\end{tablenotes}
\end{threeparttable}
\end{table}

\noindent{\textbf{Stage 4: Vulnerability Verification and Report Quality Assessment.}} In the final step, we applied two strict checks: (1) we made sure the incident involved at least one smart contract vulnerability not merely like infrastructure failure, human error, or stolen private keys; and (2) we checked that the incident was well-documented. To be included, a report had to clearly explain what the vulnerability was (including vulnerability localization information), how it was exploited, and what caused it. We excluded any cases that lacked enough technical detail.
This process yielded a final dataset of \num{50} well-documented, high-impact smart contract incidents with at least one smart contract vulnerability exploited in each incident. We use this dataset as the basis for our analysis (\Cref{sec:rq2-results}), which later resulted in a root cause-based four-tier classification of RQ3 (\Cref{sec:rq3-results}). We cite a total number of \num{126} reviewed reports for RQ2 and RQ3.

\section{Current Ethereum Smart Contract Vulnerabilities and Their Prevention/Mitigation}\label{sec:rq1-results}

To answer RQ1, we conducted a comprehensive review of the literature (\Cref{sec:lit-study-protocol}), identifying \num{29} distinct smart contract vulnerabilities. Upon further analysis, we determined that \num{5} of these vulnerabilities have become obsolete by protocol-level upgrades (\Cref{sec:deprecated-suicidal} and \Cref{sec:deprecated-call-stack-depth}), compiler improvements (\Cref{sec:deprecated-wrong-address} and \Cref{sec:deprecated-erroneoous-visibility}), or re-contextualized based on widespread usage (see \Cref{sec:deprecated-upgradable}). 
This left \activeVulnsNum\ active smart contract vulnerabilities based on the academic literature. For each of the active vulnerabilities, we collected prevention and mitigation strategies from a combination of academic sources and technical reports confirmed by expert review (\Cref{sec:active-vulnerabilities}).

\subsection{Active Vulnerabilities}\label{sec:active-vulnerabilities}

Our SLR resulted in the conclusion that a core set of \activeVulnsNum\ vulnerabilities remains exploitable and economically attractive. We unpack each of these \emph{active} threats here. For every vulnerability we (i) recap a real-world incident or a proof-of-concept vulnerability (full information is included in our companion repository), (ii) expose the coding pattern that enables it, and (iii) distill the prevention or mitigation techniques. Taken together, this section provides a current, practitioner-oriented catalog for anyone designing, reviewing, or formally verifying Ethereum smart contracts.

\subsubsection{\textbf{Integer Over/Underflow and Rounding Errors}}\label{vuln:integer-overflow}
These vulnerabilities stem from Solidity's limitations in handling of numerical data types.
\emph{Integer Over/Underflow} vulnerability arises when the result of a calculation exceeds the lower or upper limits of the variable type size, making it impossible to be expressed within that type. This is common with smaller data types. For instance, in \Cref{lst:overflow}, the balance stored in \code{balances} mapping will reset to zero if it reaches the unit value of the upper limit \(2^{256}\). Additionally, \emph{rounding errors} occur when division operations are performed on integers in Solidity, which leads to inconsistent results of calculations, especially when precision is required~\cite{torres2018osiris}.

\begin{lstlisting}[language=Solidity, caption={Integer overflow/underflow in Solidity} ,label={lst:overflow}]
contract IntegerOverflowExample {
  mapping(address => uint256) public balances;
  uint256 public constant MAX_UINT = type(uint256).max;
  function deposit(uint256 amount) public {
    require(amount > 0, "Amount must be > zero");
    balances[msg.sender] += amount; }
  function withdraw(uint256 amount) public {
     require(amount>0, "Amount must be greater than zero");
     require(balances[msg.sender]>=amount, "Insufficient");
     balances[msg.sender] -= amount; }
  function forceOverflow() public {
     balances[msg.sender]=MAX_UINT; 
     balances[msg.sender]+=1;}}
\end{lstlisting}

In the listing \ref{lst:overflow}, the function \code{forceOverflow} attempts to assign the maximum possible value to a user's balance and then increase it by one, which in earlier Solidity versions (\textless 0.8.0) would have resulted in an integer overflow. Similarly, the \code{withdraw} function reduces the user's balance, which in older versions, if unchecked, could lead to an integer underflow. Solidity 0.8.0 and later versions mitigate these issues by introducing built-in overflow and underflow checks, making such operations revert automatically.

\paragraph{\textbf{Mitigation/Prevention}}
To prevent arithmetic vulnerabilities in Solidity smart contracts, developers commonly rely on \code{SafeMath}~\cite{openzeppelin2018}, built-in checks in Solidity $0.8.0+$, and compiler warnings. When an overflow or underflow occurs, the \code{SafeMath} library reverts the transaction. 
However, developers may apply it inconsistently, and it increases gas costs. 
Solidity $0.8.0+$ eliminates the need for \code{SafeMath} by performing these checks at the compiler level, though it still introduces some runtime overhead~\cite{solidityDocs,torres2018osiris}. To prevent precision loss, developers use \code{FullMath} library~\cite{FullMathUniswap}.
Another prevention strategy in case of integer over/underflow is using suitable data types~\cite{solidityDocs}. Using formats with sufficient capacity, such as \code{uint256} instead of \code{uint8} helps prevent integer over/underflow issues.

\subsubsection{\textbf{Improper Handling of External Calls}}\label{vuln:External-Call-Failures}

Also known as ``Mishandled Exception'', ``Exception Disorder'', or ``Unchecked Call Return Value'', this vulnerability arises when a contract ignores the boolean \code{success} flag returned by a low‑level external call. In the EVM the operations \code{call}, \code{delegatecall}, \code{staticcall}, \code{callcode}, and the wrappers \code{send} and \code{transfer} never propagate a revert from the callee; instead they return a tuple \code{(success, data)}, where \code{success} is \texttt{true} on completion and \texttt{false} if the callee runs out of gas, deliberately \code{revert}s, or encounters any other error. Failure to test this flag allows the caller to proceed with an invalid state, potentially locking Ether or corrupting storage.

Two particularly common manifestations are documented in the literature~\cite{zhang2023svscanner,mense2018security}. First, when Ether is transferred with \code{send} or \code{transfer}, only a \num{2300}‑gas stipend is forwarded. Any recipient whose fallback or \code{receive} function requires more gas will revert, but the sender sees only a \texttt{false} return value. If the caller ignores that flag, the Ether remains trapped in the sending contract, sometimes irreversibly. Second, any low‑level call—whether it uses \code{call}, \code{delegatecall}, \code{staticcall}, or \code{callcode}—that proceeds without checking \code{success} suffers the same silent‑failure risk: the callee's error is suppressed, control returns to the caller, and subsequent logic executes under false assumptions.

\paragraph{\textbf{Prevention/Mitigation}}
To mitigate vulnerabilities related to external calls, developers should verify the return value of such calls. For example, using \code{require(sendResult, "Transfer failed")} after \code{send}. 
Additionally, it is recommended to use \code{transfer} instead of \code{send}, as \code{transfer} automatically reverts the entire transaction in case of failure~\cite{zhang2023svscanner}.  
Since Solidity $0.6.0$, the introduction of the \code{try/catch} structure has provided a more robust way to handle errors in external calls. Furthermore, adopting well-established standards such as \texttt{ERC20} and \texttt{ERC721}, which implement secure transaction mechanisms, is advised~\cite{solidityControlStructures}.

\subsubsection{\textbf{Reentrancy}}\label{vuln:Reentrancy}
The reentrancy vulnerability occurs when functions are maliciously reentered (e.g., through fallback functions)~\cite{zhang2023efficient}.
If the attacker is able to bypass the validity checks in the target contract, they can reenter the callee function maliciously. The vulnerability arises due to two main reasons: first, a contract's control flow decision relies on some of its state variables that should be updated by the contract itself, but are not, before calling another contract; and second, there is no gas limit when handing over the control flow to another contract. The Decentralized Autonomous Organization (DAO) attack exploited this vulnerability in \num{2016}. As shown in \Cref{fig:contracts}, the original contract includes functions for depositing and withdrawing Ether. The primary issue lies in the \code{withdraw} function, where the user's balance is set to zero after Ether has been sent to their address. This sequence allows an attacker to repeatedly call \code{withdraw} before the balance update occurs for multiple unauthorized withdrawals. In the attack contract, the attacker leverages the \code{fallback} function to invoke \code{withdraw} repeatedly before the balance is adjusted, receiving multiple illegal Ether payments.

\begin{figure}[h]
    \centering
    \begin{minipage}[t]{0.48\textwidth}
        \centering
        \lstset{abovecaptionskip=2pt, belowskip=-5pt}
        \begin{lstlisting}[language=Solidity, caption={Vulnerable Contract},label={lab:valcon}]
contract DepositFunds {
  mapping(address => uint) public balances;
  function deposit() public payable {
    balances[msg.sender] += msg.value;}
  function withdraw() public {
    uint bal = balances[msg.sender];
    require(bal > 0);
    (bool sent, ) = msg.sender.call{value: bal}("");
    require(sent, "Failed to send Ether!");
    balances[msg.sender] = 0;}}
        \end{lstlisting}
    \end{minipage}
    \hfill
    \begin{minipage}[t]{0.48\textwidth}
        \centering
        \lstset{abovecaptionskip=2pt, belowskip=-5pt}
        \begin{lstlisting}[language=Solidity, caption={Attack Contract}, label={lab:attcon}]
contract Attack {
  DepositFunds public depositFunds;
  constructor(address addr) {
    depositFunds = DepositFunds(addr);}
  fallback() external payable {
    if (address(depositFunds).balance>= 1 ether){
      depositFunds.withdraw(); }}
  function attack() external payable {
    depositFunds.deposit{value: 1 ether}();
    depositFunds.withdraw();   }}
        \end{lstlisting}
    \end{minipage}
    \hfill
    \begin{minipage}[t]{0.48\textwidth}
        \centering
        \lstset{abovecaptionskip=2pt, belowskip=-5pt}
        \begin{lstlisting}[language=Solidity, caption={Fixed Contract}, label={ lab:fixcon}]
contract SecureDepositFunds {
  mapping(address => uint) public balances;
  function deposit() public payable { balances[msg.sender] += msg.value;}
  function withdraw() public {
    uint bal = balances[msg.sender];
    require(bal > 0); balances[msg.sender] = 0;
    (bool sent, ) = msg.sender.call{value: bal}("");
    require(sent, "Failed to send!"); }}

        \end{lstlisting}
    \end{minipage}
    \caption{Vulnerable $\leftrightarrow$ attack contract pairs and the patched contract involving in reentrancy vulnerability.}
    \label{fig:contracts}
\end{figure}
 
\paragraph{\textbf{Mitigation/Prevention}}\label{sec:reentrancy-mitigation}
Two well-established techniques, Check-Effects-Interactions (CEI)~\cite{ChecksEffectsInteractions,captureDCR} and reentrancy guard locks, are commonly used to mitigate this issue. In the vulnerable \code{DepositFunds} contract (\Cref{lab:valcon}), in line 10, the balance is modified after the external call returns, enabling the malicious reentrancy. In CEI, the contract first verifies conditions (Check), then executes external interactions (Interactions), and finally updates the internal state (Effects). In the improved contract, \code{SecureDepositFunds} (\Cref{lab:fixcon}), which implements the CEI pattern, the user's balance is reset to zero before transferring the Ether. This ensures that the internal state is updated first (Effects), followed by external interactions (Interactions). By verifying the conditions initially (Check) and ensuring that no funds remain in the account before initiating the transaction, the contract prevents an attacker from making repeated withdrawals even if they attempt to exploit reentrancy~\cite{zhang2023novel}.

The second approach involves using a lock variable to prevent multiple executions of a sensitive function at the same time. When \code{withdraw} is called, the lock is activated to prevent reentry into the function until execution is complete. Once the function finishes, the lock is released, ensuring that no concurrent executions occur. This mechanism effectively neutralizes reentrancy attacks by blocking repeated calls to the function within the same transaction cycle.

\subsubsection{\textbf{Dangerous Delegatecall}}\label{vuln:Delegatecall}
\begin{figure}
    \centering
    \includegraphics[width=1\columnwidth]{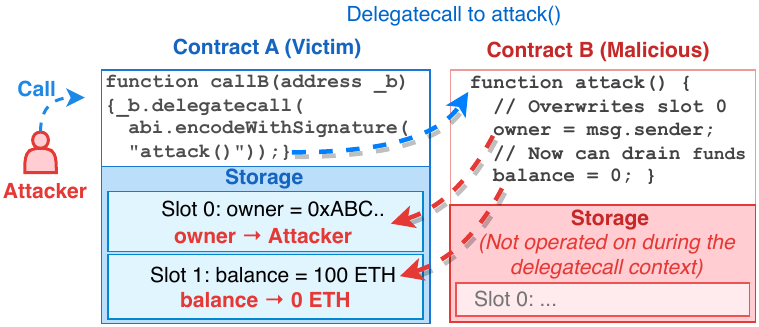}
    \caption{
    \footnotesize
    Delegatecall attack flow. A flawed interaction through a \code{delegatecall} can be initiated by the attacker calling a function in the target contract, which involves an unsafe \code{delegatecall} to an attacker-desired contract. This vulnerability might require a chain of exploits to pull off a successful attack (see \Cref{insight:multi-vuln-incidents}).}
    \label{fig:delegatecall}
\end{figure}

This vulnerability gained attention after the Parity hack in 2017, which exploited the EVM opcode \code{delegatecall}. Unlike a regular \code{call}, \code{delegatecall} executes the \emph{callee}'s bytecode while retaining the \emph{caller}'s storage, \code{msg.sender}, and \code{msg.value} (\Cref{fig:delegatecall}). Consequently, if the address provided to \code{delegatecall} points to untrusted code (code section of contract B in \Cref{fig:delegatecall}), that code runs with full write access to the caller’s state and can arbitrarily modify storage or even trigger self‑destruction, leading to loss of funds or contract functionality~\cite{jiang2018contractfuzzer,chen2020survey}.

A critical flaw was disclosed in December 2023 after several projects adopted the meta‑transaction standard ERC‑2771~\cite{ERC2771SecureProtocol}.  
Many implementations used \code{delegatecall} to let the forwarder library determine \code{\_msgSender()}, but failed to validate the forwarded payload.  
An attacker could therefore wrap arbitrary \code{calldata} inside a trusted forwarded request and trick the callee into believing that the transaction originated from any address of the attacker's choosing.  
The technique was observed in the wild, causing losses of \num{84.59} ETH and \num{17394} USDC in separate incidents~\cite{ArbitraryAddressSpoofing,UnveilingERC2771Context2023}.  

\paragraph{\textbf{Mitigation and Prevention}}
The safest way to reuse code is to rely on \emph{stateless} Solidity libraries that are linked at compile time with the \code{using for} directive; the bytecode of such a library executes in its own context, so any \code{SSTORE} or \code{SLOAD} it performs affects only the caller’s storage slot that the in‑line assembly refers to, not the library itself~\cite{soliditylibs}. When dynamic dispatch is unavoidable, restrict \code{delegatecall} to logic contracts that you have fully audited and whose bytecode hash is hard‑coded or stored behind an owner‑controlled upgrade mechanism. In practice, this means keeping the implementation address in an immutable variable or in a proxy’s \code{implementation} slot that can be changed only by a multisig with \code{onlyOwner} or \code{AccessControl}~\cite{swc112}. Never allow arbitrary user‑supplied addresses to be the target of \code{delegatecall}.
Upgradable contracts add a second risk surface: if the new implementation introduces or reorders state variables, every storage slot after the change shifts, corrupting data. To avoid this, adhere to an explicit storage‑layout discipline: reserve \emph{gaps} (\code{uint256[50] private __gap;}) at the end of each contract, append new variables only after those gaps, and use advanced Solidity code differencing tools such as \textsc{SoliDiffy}~\cite{SoliDiffy} before every upgrade to verify the intention of the modified codes. Standards that encode these patterns, notably EIP‑2535 (Diamond) and the UUPS proxy, embed storage‑slot constants and upgrade guards that make layout violations detectable on‑chain~\cite{diamondstandard,uups}. Leveraging audited reference implementations such as the OpenZeppelin \code{diamond} and \code{UUPS} modules eliminates common flaws around \code{delegatecall} and proxy upgrades~\cite{openzeppelin} (that use \code{delegatecall} to upgrade the contract logic).

\subsubsection{\textbf{ Unbounded Loops in Dynamic Arrays}} \label{vuln: Unbounded-Loops}
When a loop in a smart contract runs over all elements of a dynamic array, the gas cost for executing the loop will be directly related to the number of elements. If the array grows over time or contains many elements, the gas cost may exceed the block gas limit. This can lead to the failure of the function's execution and result in a DoS attack.

\paragraph{\textbf{Mitigation/Prevention}}
Developers should follow best practices like limiting loop iterations~\cite{ghaleb2022etainter}, dividing operations into smaller transactions, employing pull payment patterns rather than direct asset transfers in loops, and tracking gas consumption (before deployment) to improve efficiency~\cite{captureDCR,rareskills2023gas,hacken2023gas}.

\subsubsection{\textbf{DoS Attack via Owner Account}}\label{vuln: DoS-Attac}
Many smart contracts have an owner account that controls the contract. If this account is not adequately protected, attackers may exploit it, which could lead to severe consequences, such as permanently locking Ether in the contract~\cite{su2022effectively}.

\paragraph{\textbf{Mitigation/Prevention}}
Contracts should restrict access to critical functions  multi-signature approvals for transactions to avoid reliance on a single account for key operations~\cite{infuy2023dos}.

\subsubsection{\textbf{State‑Revert}}\label{vuln:State-Reverting}
A contract is vulnerable to \emph{state‑revert abuse} when it reveals the result of a state‑changing operation \emph{before} the change becomes irrevocable, allowing a caller to decide whether to let the transaction stand. The classic setting is a probabilistic reward dApp that pays one of several prizes: an attacker routes the call through an auxiliary contract, records its own balance, invokes the reward function with \code{(bool ok, bytes data) = address(dApp).call(payload)}, inspects \code{data} to see which prize was drawn, and immediately executes \code{require(false)} if the prize is unsatisfactory. Because the revert propagates upward, every write inside the dApp, including token transfers, is rolled back, yet the attacker has already learned the outcome. By repeating this cherry‑picking loop until the desirable branch appears, the adversary collects only high‑value rewards and skews the game’s expected value~\cite{liao2023smartstate}.

\paragraph{\textbf{Prevention/Mitigation}}
Effective countermeasures postpone disclosure or make it irreversible. Commit‑and‑reveal schemes write a sealed commitment (\code{keccak256(seed, user)}) in one transaction and reveal the seed in a later block, so the outcome is fixed before either party knows it. When genuine randomness is needed, verifiable random‑function services (e.g., Chainlink VRF) provide entropy that cannot be observed in time to trigger a revert. Finally, the reward function can burn gas or charge a non‑refundable fee \emph{before} emitting any tell‑tale data to eliminate the economic incentive to revert~\cite{captureDCR}.

\subsubsection{\textbf{Frozen Ether}}\label{vuln:frozen}
\emph{Frozen Ether} (a.k.a.\ ``locked money'') occurs when a contract can accept ETH but, because its withdrawal logic lives in an external library reached via \code{delegatecall}, it cannot release the funds once that library is absent. If the library address is wiped with \code{selfdestruct}—intentionally or by accident—the forwarder still receives deposits (no code is needed for that), yet every spend attempt reverts, leaving the balance permanently inaccessible~\cite{he2023detection,jiang2018contractfuzzer}.

\paragraph{\textbf{Mitigation/Prevention}}
Any contract designed to hold ETH should therefore embed an internal withdrawal function, e.g.\ using \code{call\{value: amount\}("")} or hard‑code an immutable library address that cannot be destroyed; without such a path, deposits may become irreversible~\cite{quicknodeVulnerabilities}.

\subsubsection{\textbf{Insecure Randomness}}\label{vuln:Insecure-Randomness}
This vulnerability affects contracts that require randomness (e.g., gambling and lottery contracts). These contracts often generate pseudo-random numbers using seeds such as \texttt{block.number}, \texttt{block.timestamp}, \texttt{block.difficulty}, or \texttt{blockhash}. Since these seeds can be influenced by blockchain network participants (validators or miners), malicious actors can manipulate these values to alter the outcome in their favor~\cite{reutov2018predicting,OWASP2025}. 

\begin{lstlisting}[language=Solidity, caption={Lottery Contract with insecure randomness.}, label={lst:lottery}]
contract VulnerableLottery {
  uint256 private seed;
  function play() public payable {
    require(msg.value == 1 ether);
    uint256 random = uint256(keccak256(abi.encodePacked(block.timestamp, seed)));
    if (random % 10 == 0){msg.sender.transfer(9 ether);}
seed = random;}}

\end{lstlisting}
For example, the VulnerableLottery~\cite{qian2023demystifying}  contract's dependence on \code{block.timestamp} (Listing  ~\ref{lst:lottery}) and a private \code{seed} for random number generation results in unsafe randomness. Given that miners can alter \code{block.timestamp} within a narrow range, an attacker can affect the result by sending transactions at particular times. Moreover, using a predictable \code{seed} does not generate enough entropy to allow for the prediction of future outcomes.

\paragraph{\textbf{Prevention/Mitigation}}
For secure random number generation, using blockchain-internal sources like \code{block.timestamp} and \code{block.difficulty} is insecure due to their transparency and predictability~\cite{chainlinkRandomness}. Instead, developers can utilize external oracles such as Chainlink VRF (Verifiable Random Function), which provides more tamper-resistant randomness~\cite{chainlinkVRF, ahmadjee2025decision}.  
Another approach is the commit-reveal scheme, which enhances randomness security by requiring participants to submit a concealed value first and reveal it later, preventing attackers from predicting or manipulating results~\cite{ethereumDocsCommitReveal,captureDCR}. 
Additionally, employing public-key cryptosystems like the Signidice algorithm~\cite{eos2018whitepaper}, originally developed for EOS blockchain and adopted in projects like DAO.Casino, can generate random numbers in two-party contracts. Furthermore, cross-chain oracles such as BTCRelay~\cite{btcrelay2016github} provide a trustless mechanism for Ethereum smart contracts to verify Bitcoin transactions. In order to enable the use of Bitcoin's Proof-of-Work-derived entropy as a source of randomness in Ethereum contracts, BTCRelay stores Bitcoin block headers on Ethereum and permits contracts to check Bitcoin transaction inclusion proofs. When combined with other protections against validator manipulation, such as time-delay and multi-source entropy, this strategy is useful.

\subsubsection{\textbf{Authorization Using {tx.origin}}}\label{vuln:tx.origin}

The global \code{tx.origin} holds the externally owned account (EOA) that initiated the entire transaction chain, whereas \code{msg.sender} refers to the immediate caller of a function. Using \code{tx.origin} in access control checks rather than \code{msg.sender} lets an attacker insert a malicious contract into the call flow and pass the check as the original EOA. 

In \Cref{lst:victim}, the \code{Victim.withdrawAll} function authorizes withdrawal with
\code{require(tx.origin == owner)}. The attacker contract in \Cref{lst:attacker} simply calls \code{withdrawAll}; because \code{tx.origin} still equals the owner's EOA, the transfer succeeds and forwards funds to the attacker. No owner approval occurs.

\begin{lstlisting}[language=Solidity, caption={Victim contract demonstrating the use of tx.origin}, label={lst:victim}]
contract Victim {
  address public owner;
  constructor() {owner = msg.sender;}
  function withdrawAll() public {
    require(tx.origin == owner, "Not authorized");
    payable(msg.sender).transfer(address(this).balance);
  }
receive() external payable {} }
\end{lstlisting}

\begin{lstlisting}[language=Solidity, caption={Attacker contract exploiting tx.origin vulnerability}, label={lst:attacker}]
contract Attacker {
  Victim public victimContract;
  constructor(address _victimContractAddress) {
    victimContract = Victim(_victimContractAddress);
  }
  function attack() public { victimContract.withdrawAll(); 
}  receive() external payable {} }
\end{lstlisting}

\paragraph{\textbf{Prevention/Mitigation}}
Developers should avoid using \code{tx.origin} for authorization and instead rely on \code{msg.sender}, which represents the immediate caller of the function~\cite{infuyTxOrigin}. Furthermore, it can be beneficial to restrict function access to approved users~\cite{slowmistTxOrigin} by putting strong access control mechanisms such as OpenZeppelin \code{Ownable} contract.

\subsubsection{\textbf{Front‑Running (Abusing Transaction‑Ordering Dependency)}}\label{vuln:frontRunning}
Front-running is a common attack vector that exploits the public and sequential nature of blockchain transaction processing, allowing an adversary to profit by preempting a victim's pending operation. Because miners (or validators) decide the execution order of pending transactions, an adversary who foresees a profitable state change can insert their own call ahead of the victim's. This is done either by rebroadcasting the same call with a higher gas price, which miners typically favor~\cite{chen2020survey}, or if the adversary controls the block proposer by unilaterally reordering transactions regardless of fees~\cite{zhang2024nyx}. Such re‑ordering can alter contract outcomes and extract value from honest users. In a simple scenario of front-running attack, Alice broadcasts transaction \(T_1\) with gas price \(G_1\) at time \(t_1\).
Bob observes \(T_1\) in the mempool and immediately submits \(T_2\) with a higher
gas price \(G_2 > G_1\) at \(t_2 > t_1\).
Validators, who generally sort pending transactions by gas price, include \(T_2\)
first in block \(B_x\); \(T_1\) is confirmed later (in the same or a subsequent
block).
By pre‑empting the state change encoded in \(T_1\) (e.g., a profitable DEX
swap), Bob captures its economic value, an archetypal front‑running attack that exploits transaction‑ordering dependency.

\paragraph{\textbf{Prevention/Mitigation}}
To address this vulnerability, developers can use techniques like batch auctions, which process several transactions at once to limit manipulation of transaction ordering~\cite{scsfgFrontrunning}, and commit-reveal schemes, which divide transactions into two phases to initially conceal sensitive information and reduce front-running risk~\cite{scsfgFrontrunning}. 

\subsubsection{\textbf{Prodigal Contract}}\label{sec:prodigal}
This vulnerability refers to smart contracts that wrongly send Ether to arbitrary addresses and insufficiently control Ether transfers. It is also known in literature as ``leaking ether to arbitrary addresses''. 
During runtime, sending Ether to addresses other than contract's owners or addresses that have not deposited Ether to the contract, could flag the existence of this vulnerability in a contract~\cite{nikolic2018finding}.

\paragraph{\textbf{Prevention/Mitigation}}
Thorough input validation and stringent access rules are necessary to stop prodigal contracts. Unintentional asset leaks are decreased via input validation, which makes sure parameters like recipient addresses or transfer amounts follow certain rules (such as whitelisting addresses or confirming non-zero values)~\cite{solidityDocs}. Another security measure against unauthorized activities is added by restricting important functions (e.g. fund transfers) to authorized entities through the use of Role-Based Access Control (RBAC) or OpenZeppelin's \code{onlyOwner} modifier~\cite{openzeppelin}.  

\subsubsection{\textbf{Ether Lost to Orphan Address}}\label{sec:ether_lost_orphan_addr}
To initiate an Ether transfer, a valid 160-bit address must be specified. If the provided address is invalid or does not exist, the transferred funds will be permanently lost~\cite{mense2018security}. 
If the \code{to} address is set to the contract’s own address, the transferred tokens will get locked inside the contract with no way to withdraw them, resulting in a loss. This happens because the function does not verify whether the \code{to} address is a valid recipient, including ensuring that it’s not the contract itself (see \Cref{lst:transfer}).
\begin{lstlisting}[caption={Transfer function without destination address validation}, label={lst:transfer}]
function transfer(address to, uint256 amount) public returns (bool) {
  balances[msg.sender] -= amount; balances[to] += amount;
  emit Transfer(msg.sender, to, amount); return true;
}
\end{lstlisting}

\paragraph{\textbf{Prevention/Mitigation}}
Developers should assure (via \code{require} statements) recipient addresses are neither the zero address (\code{address(0)}) nor the contract's own address (\code{address(this)}) and confirm the recipient's capacity to manage transfers (e.g., avoiding non-withdrawable contracts)~\cite{openzeppelin}. Adopting ``pull-over-push'' withdrawal mechanisms, where users initiate transfers rather than relying on direct pushes, also helps reducing risks of sending to invalid addresses~\cite{ethereum}.

\subsubsection{\textbf{Untrustworthy or Manipulable Data Feeds}}\label{vuln:Untrustworthy_Data_Feeds}

Smart contracts often depend on external facts such as token prices, weather data, and sports scores to make irreversible decisions.  
An \emph{oracle} is the component that transports those facts on chain (or acts as an service contract to provide data from other on-chain services).  
Recent studies show that integrity of the data can be manipulated at stages of its lifecycle~\cite{OracleExploitation}.  
We identified two settings where data feeds of these oracles are manipulable:

\begin{itemize}[leftmargin=*, itemsep=0pt, parsep=0pt, topsep=0pt, partopsep=0pt]
  \item \emph{Decentralized Exchange (DEX)‑based on‑chain oracles.}  
        \emph{System model:} a contract queries the current price shown by an Automated Market Maker (AMM) pool such as Uniswap~\cite{uniswap}.  
        \emph{Attacker model:} participants may inject capital (e.g., via a flash loan) to shift the pool's reserves for a single block and read back a distorted price before arbitrage restores equilibrium~\cite{DeFiRanger}.
  \item \emph{Off‑chain oracles.}
        \emph{System model:} trusted \emph{reporters} sign values off-chain and the oracle publishes them on-chain.  
        \emph{Attacker model:} compromise, bribe, or simply delay reporters to feed stale or false data~\cite{Tellor}.  
\end{itemize}

\paragraph{Prevention/Mitigation}
The literature proposes the following defense strategies:

\begin{itemize}[leftmargin=*,itemsep=1pt]
  \item \emph{Temporal aggregation.}  To deter price manipulation in DEX-based oracles, use multi‑block medians or long‑window time-weighted average pricing (TWAP) taken to smooth out single‑block distortions~\cite{OracleExploitation}.
  \item \emph{Economic deterrence.}  In case of off-chain oracles, require the reporters to put down a bond larger than the profit they could earn by lying, and slash the bond on proven fraud~\cite{Tellor,OracleExploitation}.
  \item \emph{Grace periods for finality.}  A contract should ignore data that has not yet cleared a \emph{dispute} window to block contested updates.
  \item \emph{On‑chain invariants and circuit breakers.}  Use invariants (in form of \code{require} statements) that cap per‑block price changes to stop attacks at runtime~\cite{Trace2Inv} .
  \item \emph{Real‑time monitoring.}  To detect price manipulation attacks, dynamic monitoring systems such as rule‑based (DeFiRanger~\cite{DeFiRanger}), behaviour‑model (DeFort~\cite{DeFort}), and LLM‑based (DeFiScope~\cite{DeFiScope}) detectors raise alerts, while counter‑measure agents like FlashGuard~\cite{FlashGuard} can front‑run malicious transactions.
\end{itemize}

Based on the mentioned two system settings, combining smoothing, economic incentives, program invariants, and runtime monitoring can raise the adversary's required capital, coordination, and technical difficulty beyond profitable levels.

\subsubsection{\textbf{Compiler Version Not Fixed}}\label{vuln:Compiler_Version_Not_Fixed}
This vulnerability occurs when a contract uses an outdated or broadly specified compiler version~\cite{demir2019security}. For example, specifying \code{pragma solidity ^0.4.0;} allows compilation with any version from \code{0.4.0} to \code{0.5.0} (excluding \code{0.5.0}). This makes it possible for the contract to be compiled with versions known for certain vulnerabilities. Here, in versions before \code{0.4.22}, if a function intended as a constructor did not match the contract name exactly (e.g., after renaming the contract but not the function), it was treated as a regular public function, not a constructor. This allowed attackers to invoke it maliciously after deployment. By fixing the compiler version explicitly (e.g., \code{pragma solidity 0.4.25;}), the contract becomes compilable only with \code{0.4.25}, where the \code{constructor} keyword is mandatory, thus mitigating this vulnerability~\cite{chen2020survey}.

\paragraph{\textbf{Prevention/Mitigation}}
Developers can use code linters~\cite{Ethlint} to find missing version specifications that explicitly state the compiler version (e.g., \code{pragma solidity 0.8.17;}). Furthermore, developers should review Solidity release notes prior to upgrading the project to prevent breaking changes~\cite{solidityDocs}.

\subsubsection{\textbf{Event-Ordering Bug}}\label{vuln:Event-Ordering-Bug}
If distinct sequences of transactions or function invocations lead to an unexpected behavior or final state (post-transactions), it is identified as an EO bug~\cite{kolluri2019exploiting}.

\paragraph{\textbf{Prevention/Mitigation}}
Developers should impose strict execution order by pre- and post-conditions (\code{require/assert} statements in entry and exit points of functions in Solidity) to ensure transactions follow a desired sequence to prevent event-ordering (EO) issues~\cite{captureDCR,kolluri2019exploiting}. 

\subsubsection{\textbf{Type Casting}}\label{vuln:Type-Casting}
This vulnerability stems from improper type conversions in Solidity, such as truncation bugs when casting a larger integer type (e.g., \texttt{uint16}) to a smaller one (e.g., \texttt{uint8}), leading to data loss. Additionally, converting between signed and unsigned types of the same width can cause ``signedness bugs,'' where negative values turn into large positive ones or vice versa~\cite{hwang2020gap,torres2018osiris}. 

\paragraph{\textbf{Prevention/Mitigation}}
To prevent type-casting vulnerabilities, developers should ensure proper type validation by explicitly checking types and validating value ranges before performing conversions. This reduces the risk of unintended truncation~\cite{ethereumDocsTypeCasting}. Additionally, Solidity requires explicit casting to help reduce the risk of implicit conversions, especially between signed and unsigned numbers, which might introduce unpredictable behavior ~\cite{solidityDocsGuide}. Refactoring the project to the most recent Solidity version is also beneficial as newer versions support type safety features~\cite{solidityLatestVersion}.

\subsubsection{\textbf{Ponzi Scheme}}\label{sec:ponzi}
In Ponzi contracts, vulnerabilities arise (for the users of the Ponzi services) due to low transparency and deceptive promises of high returns, and legal noncompliance. These schemes commonly rely on a constant influx of new investors to create an illusion of profitability, with returns paid to earlier participants using the funds contributed by newer ones, rather than from actual profits. This scheme is unsustainable and leads to losses for later investors when the inflow of new investments slows or stops~\cite{zheng2023securing}.

\paragraph{\textbf{Prevention/Mitigation}}
User education is important and users must avoid them by spotting characteristics of Ponzi schemes, such as unsustainable returns and a lack of transparency (unavailable, unverified, or obscured contract source code)~\cite{investopediaPonzi}. 

\subsubsection{\textbf{Storage Collision}}\label{vuln:Storage-Collision}
Storage Collision occurs when two contracts accessing the same storage space have a different understanding of the storage layout~\cite{NotYourTypeNDSS,solidity2025,wustholz2020harvey,SWC_124}. This vulnerability is often observed when a proxy upgradability design pattern is used~\cite{ProxyUpgradePattern}. 
The proxy upgradability design pattern separates a contract’s state from its logic, allowing the system’s functionality to evolve without losing stored data~\cite{captureDCR}. In this design, a proxy contract holds the persistent state and delegates calls to an external implementation contract using the \code{delegatecall} opcode. Initially, the proxy directs calls to a basic implementation. Later, when new features or patches are required, a new implementation contract is deployed. By updating the proxy's reference to this new contract, the upgraded logic is activated while the original state (the users’ balances, etc.) remains intact~\cite{ProxyUpgradePattern,captureDCR}. 
A specific subcategory of this vulnerability is also known as \emph{type confusion}. This specific subcategory occurs when the runtime context of the logic contract (e.g., in an upgradable contract) confuses the type of an object in storage with another type, leading to its incorrect utilization. For example, when a variable is allocated with one type and later accessed with an incompatible type~\cite{NotYourTypeNDSS,yao2024scif}.

For example, the smart contract setup in \Cref{lst:collision} is vulnerable to {storage collision} due to the improper alignment of storage slots between the \code{Proxy} and \code{Logic} contracts when using \code{delegatecall}. Specifically, both contracts store critical state variables in \emph{slot 0x0}, but with different interpretations: the \code{Proxy} contract uses it for \code{visits}, while the \code{Logic} contract stores \code{initialized} and \code{admin} in the same slot. When \code{initialize()} is called via \code{delegatecall}, it writes \code{true} (1) to \code{initialized} and sets \code{admin} to \code{msg.sender}, unknowingly overwriting the \code{visits} variable in the \code{Proxy} contract. An attacker can exploit this by re-invoking \code{initialize()}, since the overwritten storage allows it to be called again, setting their own address as \code{admin}. Once this is done, the attacker can execute \code{withdraw()} to drain the contract's funds. See \Cref{lst:collision} for more details.

\begin{lstlisting}[caption={Wallet Contract with storage collision vulnerability}, label={lst:collision}, language=Solidity]
contract Proxy {
  uint public visits; // Storage slot [0x0]
  address public LOGIC; // Storage slot [ERC-1967]
  constructor(address logicAddress) {
    LOGIC = logicAddress;
    (bool success, ) = LOGIC.delegatecall(
        abi.encodeWithSignature("initialize()"));
    require(success, "Initialization failed");}
  fallback() external payable {
    (bool success, ) = LOGIC.delegatecall(msg.data);
    require(success, "Delegatecall failed");}}
contract Logic {
  bool public initialized; // Storage slot [0x0]
  address public admin; // Storage slot [0x0] (Collision issue)
  uint[] public artworkIDs; // Storage slot [0x1]
  mapping(address => uint) public artworkHolders; // Storage slot [0x2]
  function initialize() external {
    require(!initialized, "Already initialized");
    initialized = true;
    admin = msg.sender; // Overwrites Proxy's storage slot [0x0] }
  function withdraw() external {
    require(msg.sender == admin, "Not admin");
    payable(admin).transfer(address(this).balance);}}
\end{lstlisting}

\paragraph{\textbf{Prevention/Mitigation}}
To avoid storage collisions in proxy patterns, keep the implementation’s storage layout consistent with the proxy’s reserved slots (or assign fixed slots with inline assembly, as in EIP‑1967)~\cite{chainlightStorageLayouts}. Supplement this with exhaustive tests and fuzzers like Harvey to catch any remaining overlaps pre‑deployment~\cite{wustholz2020harvey}.

\subsubsection{\textbf{Business Logic Flaws}}\label{vuln:Business_logic_errors}
In smart contracts, business logic flaws (aka accounting errors) occur when financial calculations are inconsistent, inaccurate, or disorganized because of type errors, misinterpreted or mismanaged units ({token units}), or the implementation is inconsistent with expected behavior by the protocol designer~\cite{zhang2024towards}. 

One such vulnerability happened in \code{Vader} protocol~\cite{Web3Bugs16}, where a token unit mismatch occurs when token values are incorrectly combined without the appropriate unit conversion, resulting in an error in the liquidity calculation (\code{calcLiquidityUnits} in \Cref{lst:pools}). In this case, the scales of the \code{base} and \code{token} tokens differ. Nevertheless, they are added together in the final formula without the proper conversion, which causes users to lose money and receive an inaccurate liquidity (\Cref{lst:pools}).

\begin{lstlisting}[language=Solidity, caption=Smart Contract: Pools, label=lst:pools]
contract Pools {
  function addLiquidity(address base, address token, address member) 
    external returns (uint liquidity) {
    uint addedBase = getAddedAmount(base, ...);
    uint addedToken = getAddedAmount(token, ...);
    liquidity = calcLiquidityUnits(addedBase, totalBase, addedToken, totalToken, totalLiquidity);
    liquidity [...][member] += liquidity;
    totalBase += addedBase; }
  function calcLiquidityUnits(uint b, uint B, uint t, uint T, uint P) 
    external view returns (uint) {
    uint part1 = (t * B);
    uint part2 = (T * b);
    uint part3 = (T * B) * 2;
    uint _units = (((P * part1) + part2) / part3);
    return (_units) / one; }}
\end{lstlisting}

\paragraph{\textbf{Prevention/Mitigation}}
Developers should write extensive test cases covering important business logic scenarios to validate code behavior. This practice helps in identifying discrepancies between the intended and actual functionalities of the smart contract. Extensive testing using model-based testing can deter business logic vulnerabilities~\cite{XploGen,ModCon}.
Implementing strict input validation ensures that only correctly formatted and expected data is processed by the contract to prevent unintended behaviors arising from malicious inputs~\cite{OWASP2025}.

\subsubsection{\textbf{Dangerous Balance Inequality}}\label{sec:dangerous_balance_inequality}
Input validation and invariant preservation are checked/enforced using \code{assert} and \code{require} statements in Solidity. Solidity smart contracts on Ethereum alone include more than \num{4.5} of these unique checks~\cite{DISL}. In such condition checks, using ``\verb|==|'' instead of ``$\geq$'' (in statements such as $a \geq b$) 
when verifying inventory arises from the flawed assumption that the exact match will always occur. 
This practice can lead to a vulnerability, as even minor, unexpected changes in inventory values, whether intentional or accidental, could bypass verification logic~\cite{demir2019security}. 

\paragraph{\textbf{Prevention/Mitigation}}
Developers should use ``$\geq$'' to account for potential pre-existing balances~\cite{solidityAssert}. Slither~\cite{feist2019slither}, the static analysis tool, provides a ``dangerous-strict-equality'' detector to identify this vulnerability. Moreover, Continuous Integration (CI) pipelines can automatically check the semantic strength of a pre-/post-condition check in Solidity using \textsc{SInDi}~\cite{SInDi}.

{
\setlength{\tabcolsep}{3pt}
\renewcommand{\arraystretch}{1.1}

\begin{table*}[!t]
\centering
\footnotesize
\caption{Summary of deprecated smart contract vulnerabilities and their mitigation}
\label{tab:deprecated-vulnerabilities}
\begin{tabular}{p{2.7cm} p{0.8cm} p{14cm}}
\toprule
\textbf{Vulnerability} & \textbf{Sec.} & \textbf{Deprecation Reason(s)} \\
\midrule

Upgradable Contracts & \S\ref{sec:deprecated-upgradable} & Re-contextualized from a vulnerability to a technical feature. Its risks are manageable through robust on-chain governance and verifiable evolution patterns. \\
Wrong Address & \S\ref{sec:deprecated-wrong-address} & Mitigated by ABI-decoding requirements in Solidity v0.5.0. Calls with \code{calldata} shorter than the expected arguments will revert, preventing the padding that enabled the attack. \\

Erroneous Visibility & \S\ref{sec:deprecated-erroneoous-visibility} & Addressed by Solidity v0.5.0, which mandated explicit function visibility declarations (\texttt{public}, \texttt{private}, etc.).
\\
Suicidal Contract & \S\ref{sec:deprecated-suicidal} & The behavior of the \texttt{selfdestruct} opcode was severely curtailed by EIP-6780. It now only functions if executed in the same transaction as contract creation, neutralizing its use in common attack patterns. \\

Call-stack Depth Limit & \S\ref{sec:deprecated-call-stack-depth} & Rendered economically infeasible by the Tangerine Whistle hard fork (EIP-150)~\cite{eip150}, which introduced the ``63/64 gas rule.'' This rule limits the gas forwarded in a call, preventing deep recursion attacks. \\
\bottomrule
\end{tabular}
\end{table*}
}

\subsubsection{\textbf{Resource Exhaustion}}\label{vuln:Resource_Exhaustion}
The gas mechanism in EVM is used to determine how much it costs to execute smart contracts. This system should guarantee that each instruction's execution cost is calculated according to the resources it uses, including CPU, RAM, and I/O. Nevertheless, some processes, such as \code{blockhash} and \code{balance}, which require access to blockchain-stored data, are substantially underpriced compared to their actual computational and storage impact due to inconsistent gas pricing for specific EVM instructions~\cite{BrokenMetre}.   
One practical example is looping through a list of accounts to repeatedly retrieve their \code{balance}.
Since the gas cost for this operation is lower than its actual computational overhead, an attacker can craft transactions that continuously invoke this instruction, causing a significant workload on the network with \emph{minimal gas expenditure for the attacker}.
This shows how inconsistencies in EVM's resource metering (vs opcode gas) can be leveraged to create transactions that disproportionately consume computational power.

\paragraph{\textbf{Prevention/Mitigation}}
Ethereum clients have undergone several upgrades to mitigate resource exhaustion risks. In May 2024, a flaw was exposed in the \emph{Geth} client's \texttt{FeeHistory} interface that allowed \texttt{RPC} requests with up to \num{600}k \texttt{rewardPercentiles} due to missing limits, leading to excessive memory usage and node crashes. The issue was resolved by capping the number of \texttt{rewardPercentiles}~\cite{SnykVulnerabilityDatabaseREx,GETHIssue}.

\subsubsection{\textbf{Access Control}} \label{vuln:Access_Control}
An access control vulnerability occurs when a smart contract improperly manages access to critical functionalities, allowing unauthorized users to carry out operations including asset transfers, ownership changes, and significant contract settings modifications. Usually, this vulnerability results from the lack of a suitable authentication method. Usual ways to authenticate include implementing Role-Based Access Control (RBAC)~\cite{PermissionBug} using \code{require} statements at the entry point of the function body or using a Solidity \code{modifier}. This vulnerability is also frequently caused by defining functions as \code{public} when only the contract owner or authorized users should be able to access them, failing to implement access control conditions like \code{require(msg.sender == admin)} and exposing sensitive variables like \code{owner} or \code{balance} as \code{public} when they should be \code{private} to prevent unwanted access~\cite{nguyen2022mando,rossini2023use,wang2022gvd}.
For example, because there is no \code{require(msg.sender == owner, "Not authorized")} check in \Cref{lst:accessControl} (after line \num{5}), any user can alter the contract owner using the public function \code{changeOwner}. Likewise, the \code{withdraw} function has no access restriction, so anyone can take money out without being authenticated.
\begin{lstlisting}[language=Solidity, caption={VulnerableContract Example}, label={lst:accessControl}]
contract AccessControlVulnerableContract {
  address public owner; uint256 public funds;
  constructor() { owner = msg.sender;}
  function changeOwner(address _newOwner) public { owner = _newOwner;}
  function withdraw(uint256 _amount) public {
    require(_amount <= funds, "Insufficient funds");
funds -= _amount; payable(msg.sender).transfer(_amount); }}  
\end{lstlisting}

\paragraph{\textbf{Prevention/Mitigation}}
Developers can utilize OpenZeppelin's \code{Ownable} and \code{AccessControl} contracts, which offer role management, to avoid access control vulnerabilities.  The \code{onlyOwner} modifier~\cite{openzeppelinOwnable} in the \code{Ownable} contract is used to limit sensitive operations to the \code{owner} role. \code{AccessControl} in this library enables defining roles and their permissions~\cite{openzeppelinAccessControl}. 

\subsubsection{\textbf{Timestamp Dependence}}\label{vuln:Timestamp_Dependence}
This vulnerability occurs when the contract's logic depends on the value of \code{block.timestamp}. Due to the clock drift of distributed systems, miners are allowed to adjust the timestamp within a specific range, typically \(\pm 900\) seconds in Ethereum~\cite{halborn2023}. Smart contract condition checks may be impacted by this manipulation, even if they have nothing to do with producing random numbers. For instance, consensus participants can alter the end time of events like bids or auctions~\cite{getsecureworld2023}. 
Validators can lengthen waiting or lock-up periods before specific activities to change the outcome to the advantage of a particular group~\cite{captureDCR}. Additionally, it is a typical manipulation for miners to reorganize and modify the timing of transactions, particularly in contracts that depend on the precise timing or order of transactions. Changing the timestamp may also alter the outcome of computations that depend on time, causing the contract to behave inconsistently. Miners may modify the voting schedule and outcomes to benefit themselves or a certain group, which could have an effect on governance ecosystems~\cite{forumopenzeppelin2023}.

\paragraph{\textbf{Prevention/Mitigation}}
Developers should use trusted time oracles to ensure accurate data sources~\cite{GloryPraise2023,cobalt2023}.

\subsection{Deprecated Vulnerabilities}\label{sec:deprecated-vulnerabilities}

\begin{table*}[!t]
\caption{Overview of studied security incidents including attack dates, financial losses in USD, and identified vulnerabilities.}
\label{tab:main_attack_attributes}
\centering
\scriptsize
\begin{tabular}{@{} l p{3cm} c r l l p{5.2cm} p{0.9cm} @{}}
\toprule
\textbf{ID} & \textbf{Incident} & \textbf{Date} & \textbf{Loss (USD)} & \textbf{Ecosystem} & \textbf{Application} & \textbf{Vulnerability} & \textbf{Network} \\
\midrule

\hypertarget{row1}{}1 & Euler Fin.~\cite{immunebytes2023euler,rekteulerrektzh,slowmist2023euler}  & 2023-03-13 & \num{197000000} & DeFi & Lending & Access control, Business Logic Flaws & Mainnet \\
\hypertarget{row2}{}2 & Nomad Bridge~\cite{nomad_rekt_2022, nomad2022} & 2022-08-02 & \num{190000000} & Bridge & Cross-chain & Access Control & Mainnet \\
\hypertarget{row3}{}3 & BonqDAO~\cite{rektNews} & 2023-02-03 & \num{120000000} & DeFi & Stablecoin & Price Manipulation & Sidechains \\
\hypertarget{row4}{}4 & WooFi~\cite{rekt_woofi_2024, olympix2023woofi} & 2024-03-06 & \num{85000000} & DeFi & DEX & Price Manipulation & L2 \\
\hypertarget{row5}{}5& Fei Rari ~\cite{rektnews2022, certik2022fei}& 2022-05-01 & \num{80000000} & DeFi & Lending & Reentrancy & Mainnet \\
\hypertarget{row6}{}6 & Infini~\cite{infini2025, yona2025rogue} & 2025-02-24 & \num{49500000} & DeFi & Stablecoin & Access Control & Mainnet \\
\hypertarget{row7}{}7 & KyberSwap~\cite{rektnews2023,slowmist2023}& 2023-11-23 & \num{48000000} & DeFi & DEX & Business Logic & Mainnet \\
\hypertarget{row8}{}8 & Penpie~\cite{olympix2024,solidityscan2024} & 2024-09-03 & \num{27000000} & DeFi & Yield Aggregator & Reentrancy & Mainnet \\
\hypertarget{row9}{}9 & Sonne Finance~\cite{certik2024, rekt2023} & 2024-05-16 & \num{20000000} & DeFi & Lending  & Rounding Error & L2 \\
\hypertarget{row10}{}10 & Inverse Fin.\cite{rekt2022, inversefinancebinance,oklink2022inverse} & 2022-04-02 & \num{15600000} & DeFi & Lending & Price Manipulation & Mainnet \\
\hypertarget{row11}{}11 & Holograph~\cite{blockbasis2024,halborn2024} & 2024-06-13 & \num{14400000} & Token & NFT  & Access Control & Mainnet \\
\hypertarget{row12}{}12 & Deus DAO ~\cite{quillaudits2023,cyvers2023} & 2023-05-05 & \num{6500000} & DeFi & Token &  Access Control & L2 \\
\hypertarget{row13}{}13 & Ionic~\cite{linkedin2025,ionicmoneyrekt} & 2025-02-04 & \num{6900000} & DeFi & Lending & Input Validation & L2 \\
\hypertarget{row14}{}14 & Ronin Bridge~\cite{roninrektii, roninbridgeanalysis} & 2024-08-06 & \num{12000000} & Bridge & Cross-chain Bridge & Access Control & Mainnet \\
\hypertarget{row15}{}15 & LI.FI~\cite{li_fi_incident_2023, solidityscan_li_fi_analysis_2023, slowmist_july_report_2023} & 2024-07-16 & \num{11600000} & DeFi & Cross-chain Bridge & Access Control & Mainnet \\
\hypertarget{row16}{}16 & Yearn Finance~\cite{rekt2023yearn, certik2023yearn} & 2023-04-13 & \num{11400000} & DeFi & Yield Aggregator & Misconfiguration & Mainnet \\
\hypertarget{row17}{}17 & zkLend~\cite{slowmist2025zklend, zklend2025linkedin} & 2025-02-12 & \num{9600000} & DeFi & Lending & Rounding Error & L2 \\
\hypertarget{row18}{}18 & LI.FI Protocol\cite{lifi2023incident, rekt2023lifi} & 2024-07-16 & \num{11600000} & DeFi & Cross-Chain   & Unchecked Call Return Value, Access Control & Mainnet \\
\hypertarget{row19}{}19 & Hundred Finance~\cite{rekt2023hundred, cointelegraph2024hundred} & 2023-04-15 & \num{7400000} & DeFi & Lending & Exchange Rate Manipulation, Rounding Error & L2 \\
\hypertarget{row20}{}20 & Abracadabra Money\cite{rekt2024abracadabra} & 2024-01-31 & \num{6500000} & DeFi & Lending & Business Logic, Rounding Error & Mainnet \\
\hypertarget{row21}{}21 & Lodestar Finance~\cite{certikLodestar2022, lodestarRekt2022} & 2022-12-10 & \num{6500000} & DeFi & Lending & Price Manipulation & L2 \\
\hypertarget{row22}{}22 & 1Inch~\cite{rekt_1inch_2024, decurity_1inch_postmortem_2024} & 2025-03-06 & \num{5000000} & DeFi & DEX & Integer Underflow & Mainnet \\
\hypertarget{row23}{}23 & Shezmu~\cite{blockapex_shezmu_2024, cyberstrategy1_truths_2024} & 2024-09-20 & \num{4900000} & DeFi & Lending & Access Control & Mainnet \\
\hypertarget{row24}{}24 & Gamma Strategies\cite{rekt2024gamma, verichains2024gamma}& 2024-01-04 & \num{4500000} & DeFi & Liquidity Pool & Price Manipulation & L2 \\
\hypertarget{row25}{}25 & Conic Finance~\cite{conic2023rekt,gamma2024analysis} & 2023-07-25 & \num{4200000} & DeFi & Lending & Read-Only Reentrancy, Price Manipulation & Mainnet \\
\hypertarget{row26}{}26 & KiloEx~\cite{halborn2025kiloex, rekt2025kiloex}& 2025-04-17 & \num{7500000} & DeFi & DEX & Access Control, Price Manipulation & L2 \\
\hypertarget{row27}{}27 & SIR Trading~\cite{cointelegraph2025sirtrading, rekt2025sirtrading}& 2025-03-30 & \num{355000} & DeFi & Synthetics &  Storage Collision & Mainnet \\
\hypertarget{row28}{}28 & Abracadabra Money\cite{halborn2025abracadabra} & 2025-03-26 & \num{13000000} & DeFi & Lending & Business Logic & L2 \\
\hypertarget{row29}{}29 & DeltaPrime ~\cite{rekt2024deltaprime, solidityscan2024deltaprime} & 2024-11-11 & \num{4850000} & DeFi &   Lending & Improper Input Validation & L2 \\
\hypertarget{row30}{}30 & Onyx Protocol~\cite{halborn_onyx_protocol_hack, onyx_protocol_rekt2}& 2024-09-26 & \num{3800000} & DeFi & Lending &Exchange Rate Manipulation, Access Control & Mainnet \\
\hypertarget{row31}{}31 & Dexible~\cite{rekt2023dexible, shashank2023dexible} &2023-02-20 & \num{2000000} & DeFi & DEX Aggregator & Access Control & Mainnet \\
\hypertarget{row32}{}32 & Dough Finance~\cite{dehacker2024dough,certik2024dough} & 2024-07-12 & \num{2100000} & DeFi & Lending & Improper Input Validation & Mainnet \\
\hypertarget{row33}{}33 & CloberDEX~\cite{cloberdex_solidityscan_2024,cloberdex_rekt_2024} & 2024-12-10 & \num{501279} & DeFi & DEX & Reentrancy & L2 \\
\hypertarget{row34}{}34 & Tornado Cash~\cite{rekt2023tornado,halborn2023tornado} & 2023-05-20 & \num{2173500} & DeFi & Privacy Mixer & Governance & Mainnet \\
\hypertarget{row35}{}35 & Team Finance~\cite{rekt2023teamfinance,slowmist2023teamfinance} & 2022-10-27 & \num{15800000} & DeFi & Token Locker & Input Validation & Mainnet \\
\hypertarget{row36}{}36 & Raft Protocol~\cite{raft-rekt,sharkteam-analysis-raft} & 2023-11-10 & \num{3600000} & DeFi & Stablecoin & Rounding Error & Mainnet \\
\hypertarget{row37}{}37 & Rho Market~\cite{rekt2024rho, olympix2024rho} & 2024-07-19 & \num{7500000} & DeFi & Lending &  Priviledge Abuse, Price Manipulation & L2 \\
\hypertarget{row38}{}38 & UwuLend~\cite{rekt2024uwulend, shashank2024uwulend} & 2024-06-10 & \num{19400000} & DeFi & Lending &  Price Manipulation & Mainnet \\
\hypertarget{row39}{}39 & Velocore~\cite{rekt2024velocore, velocore2024postmortem} & 2024-06-02 & \num{6800000} & DeFi & AMM & Underflow & L2 \\
\hypertarget{row40}{}40 & Curio~\cite{rekt2024curio, behnke2024curio} & 2024-03-23 & \num{16000000} & DeFi & DAO voting system & Access Control, Dangerous Delegatecall & Mainnet \\
\hypertarget{row41}{}41 & Unizen~\cite{rekt2024unizen, blockbasis2024unizen} & 2024-03-08 & \num{2100000} & DeFi & DEX & Dangerous Delegatecall & Mainnet \\
\hypertarget{row42}{}42 & Seneca Protocol~\cite{rekt2024seneca, blockapex2024seneca} & 2024-02-28 & \num{6400000} & DeFi & CDP & Input Validation & Mainnet \\
\hypertarget{row43}{}43 & OKX DEX~\cite{rekt2023okxdex, olympix2023okxattack} & 2023-12-12 & \num{2700000} & DeFi & DEX & Access Control, Dangerous delegatecall & Mainnet \\
\hypertarget{row44}{}44 & Zunami Protocol~\cite{rekt2023zunami, solidityscan2023zunami} & 2023-08-13 & \num{2100000} & DeFi & Stablecoin & Price Manipulation & Mainnet \\
\hypertarget{row45}{}45 & EraLend~\cite{rekt2023eralend, cryptorank2023eralend} & 2023-07-25 & \num{3400000} & DeFi & Lending & Read-Only Reentrancy, Price Manipulation & L2 \\
\hypertarget{row46}{}46 & Atlantis Loans~\cite{rekt2023atlantis, solidityscan2023atlantis} & 2023-06-10 & \num{2500000} & DeFi & Lending & Governance & Sidechains \\
\hypertarget{row47}{}47 & Sturdy Finance~\cite{rekt2023sturdy, solidityscan2023sturdy} & 2023-06-12 & \num{800000} & DeFi & Lending & Read-only Reentrancy, Price Manipulation & Mainnet \\
\hypertarget{row48}{}48 & Jimbo's Protocol~\cite{rekt_news_jimbo, numen_cyber_labs_jimbo} & 2023-05-28 & \num{7500000} & DeFi & Lending/DEX & Access Control, Price Manipulation & L2 \\
\hypertarget{row49}{}49 & Swaprum~\cite{rekt2023swaprum, cryptorank2023swaprum} & 2023-05-19 & \num{3000000} & DeFi & DEX & Access Control & L2 \\
\hypertarget{row50}{}50 & Orion Protocol~\cite{rekt2023orion, slowmist2023orion} & 2023-02-03 & \num{3000000} & DeFi & DEX Aggregator & Reentrancy & Mainnet \\

\bottomrule
\end{tabular}
\end{table*}

This section discusses vulnerabilities in smart contracts that were formerly seen as security issues but have been fixed by protocol enhancements and governance systems. \Cref{tab:deprecated-vulnerabilities} presents a list of these deprecated Ethereum smart contract vulnerabilities and their brief deprecation reasons.

\subsubsection{\textbf{Upgradable Contracts}}\label{sec:deprecated-upgradable}
 
Smart contracts often incorporate upgradability mechanisms through logic contracts and proxies to manage immutability constraints of the code section of the contract~\cite{wood2014ethereum}. Although early analyses classified upgradability as a potential vulnerability~\cite{chen2020survey}, recen advancements in upgradability patterns and governance mechanisms have demonstrated that this is better understood as a technical challenge rather than a security weakness. 
On-chain governance proposals~\cite{compound_governance} now enable decentralized control over upgrades, reducing the risk of unilateral malicious actions. Moreover, research efforts such as SoliNomic~\cite{SoliNomic} have introduced formal refinement-based approaches to structuring the evolution of decentralized organizations (for correct and consistent upgrades). Therefore, upgradability, when properly designed, enhances the adaptability of DApps without compromising security.

\subsubsection{\textbf{Wrong Address}}\label{sec:deprecated-wrong-address}
This vulnerability arises due to the improper handling of function arguments. EVM expects input arguments to be encoded in chunks of 32 bytes; however, if an attacker deliberately sends a truncated address (i.e., an address with missing bytes at the end), EVM automatically pads the missing bytes with zeros, potentially altering the intended values of other arguments. This misalignment can cause an attacker to manipulate transaction amounts or other critical parameters in smart contracts~\cite {he2023detection,chen2020survey}. 
The attacker could craft a transaction with a shortened address, which, when processed, causes the EVM to pad the missing bytes with zeros, leading to a shift in memory alignment where other function arguments, such as transfer amounts, are misinterpreted (e.g., excessive transfers to the attacker's address). 
To prevent this vulnerability, developers should implement strict validation to check that the input data matches the expected size. In other words, the contract should explicitly check \code{msg.data} length to ensure proper alignment of the arguments.
In modern Solidity versions ($\geq$0.5.0), with its standard ABI decoding, the short address attack is prevented by design~\cite{solidity_v050_breaking_changes}. In these modern versions, if an attacker tries to supply a truncated address (resulting in insufficient \code{calldata}), the function call will revert before any parameter decoding or auto-padding occurs. 

\subsubsection{\textbf{Erroneous Visibility}}\label{sec:deprecated-erroneoous-visibility}
Erroneous visibility vulnerability emerged from Solidity's default function visibility behavior prior to version \texttt{0.5.0}. Functions without explicit visibility modifiers automatically became public, exposing critical operations to unauthorized external calls~\cite{piantadosi2023detecting}. Implicit behavior often led to unintentional access to state-modifying, fund management, or administrative functionality. Missing visibility specifiers allowed attackers to execute privileged functions in the Parity multisig wallet incident~\cite{tsankov2018securify}.

By requiring explicit visibility declarations for all functions, Solidity version \code{0.5.0} released in November 2018, fixed this vulnerability~\cite{solidity050breaking}. Contracts without visibility specifiers are now rejected by the compiler, turning a runtime vulnerability into a compile-time error. \code{public}, \code{private}, \code{internal}, or \code{external} visibility must be explicitly declared by functions.

\subsubsection{\textbf{Suicidal Contract}}\label{sec:deprecated-suicidal}
EVM's \code{selfdestruct} allows the contract to destroy itself and send the remaining Ether in the contract to a target address.
Since the \emph{Dencun} upgrade (EIP‑6780, March 2024) \code{selfdestruct} no longer removes code or storage and no longer grants a gas refund, \emph{unless it is invoked in the same transaction that deployed the contract}~\cite{eip6780}. Outside that narrow case, it merely transfers the contract's Ether balance to the specified address, effectively disabling contract ``suicide'' and the attacks it enabled.

\subsubsection{\textbf{Call‑Stack Depth Limit}}\label{sec:deprecated-call-stack-depth}
The EVM caps a call chain at \(1024\) frames.  
Before EIP‑150~\cite{eip150} an attacker could, for about \num{40000} gas, create \(1024\) nested calls and cause every subsequent \code{call} from a victim contract to fail (return 0).  
EIP‑150 raised the sub‑call base cost to \num{700} gas and, more importantly, restricted each call to forwarding at most \(63/64\) of its remaining gas. Because the available gas shrinks geometrically, filling the stack now exceeds the block gas limit, rendering the ``stack‑depth attack'' economically infeasible.  
The depth cap nonetheless remains, so contracts should always check the return value (or use \code{try/catch}) when interacting with external contracts.

\section{How Do Real-World Attacks on Ethereum Smart Contracts Align With Academic Vulnerability Classifications?}\label{sec:rq2-results}

To answer RQ2, we conducted a comprehensive analysis of \num{50} significant smart contract exploits that occurred between \num{2022} and early \num{2025}. The collection, selection, and analysis protocol for these attacks is presented in \Cref{sec:incident-review-protocol}.
\Cref{tab:main_attack_attributes} presents the main attributes of these attacks, including date, volume of financial loss, ecosystem, application type, and the exploited vulnerabilities (identified through manual expert analysis and based on active vulnerabilities in \Cref{sec:active-vulnerabilities}). We present more information of each attack and its attributes (victim contact and attacker addresses, etc.) along with a summary of the technical details of each attack in our accompanying repository.

Our insights reveal that while real-world successful attacks involve technical weaknesses attributed to classes of known vulnerabilities (\Cref{sec:active-vulnerabilities}), there are dominant patterns in the incidents that have been unnoticed by the literature. Access control vulnerabilities and price manipulation emerged as the most frequent and financially devastating attack vectors, accounting for a combined loss of \$\num{697.7} million in our dataset (see \Cref{tab:vulnerability_distribution}). 
Access control-related incidents involve exploit methods beyond mere coding flaws (e.g., lack or misplacement of \code{require(msg.sender==owner)} checks) and include human-centric factors as well (see \insightref{insight:human-errors}).  
The high frequency of price manipulation attacks underscores the critical importance of secure, manipulation-resistant oracles (\insightref{insight:oracle}). We also observed that while classic vulnerabilities like reentrancy persist, attackers are employing more sophisticated variants like \emph{read-only reentrancy} and targeting new EVM features such as transient storage (\insightref{insight:new-evm-features}), opening up novel exploit avenues.

\paragraph{Incidents are rarely single-shot failures}
Only 19 of the \num{50} exploits hinged on merely a code defect; the rest were \emph{chains} of two or more weaknesses (\insightref{insight:multi-vuln-incidents}) or contained a human factor enabling the incident.  For instance, governance mistakes (on a protocol design level) served as the entry point, which attackers then combined with a price-feed or arithmetic flaw to drain funds.

\paragraph{Human factors are involved}
Roughly one-third of the breaches trace back to mis-set upgrade parameters, retained developer keys, or copy-paste mistakes-issues that automated analyses seldom flag (\insightref{insight:human-errors}).  Where the literature focuses on the correctness of implementation, our data show that operational discipline and key management are at least as consequential.

\paragraph{Amplifiers and emergent surfaces}
Flash-loan liquidity reliably magnified price manipulation attacks (\Cref{insight:flash}); protocol composability let a rounding bug in one contract propagate across lending markets (\insightref{insight:composability}); and brand-new EVM features such as transient storage introduced fresh attack vectors (\insightref{insight:new-evm-features}). These observations motivate the tiered root-cause model introduced later in \Cref{sec:rq3-results}. 

Insights 1--10 present the identified patterns in detail.

\begin{table}[!t]
\centering
\caption{Breakdown of security vulnerabilities by frequency and financial impact across analyzed incidents}
\label{tab:vulnerability_distribution}
\begin{tabular}{ l c r }
\hline
\textbf{Vulnerability Category} & \textbf{Number of Incidents} & \textbf{Total Losses (USD)}  \\
\hline
Access Control & 13 & 417,950,000  \\
\hline
Price Manipulation & 13 & 279,750,000 \\
\hline
Reentrancy & 7 & 118,501,279 \\
\hline
Business Logic Flaws & 5 &  177,400,000  \\
\hline
Input Validation & 6 & 41,850,000 \\
\hline
Rounding Error & 4 & 36,900,000  \\
\hline
Integer Underflow & 2 &  11,800,000  \\
\hline
Governance & 2 &  4,673,500 \\
\hline
Dangerous Delegatecall & 2 & 4,800,000  \\
\hline
Storage Collisions & 1 &  355,000  \\
\hline
\end{tabular}
\end{table}

\begin{InsightBox}[{insight:human-errors}]{The Pervasiveness of Operational Human Mistakes}
Human configuration errors manifest in ways such as: (1) \emph{deployment/upgrade flaws:} we witnessed misconfigurations during upgrades (Nomad Bridge~\cite{nomad_rekt_2022}), (2) incomplete upgrades (Ronin Bridge~\cite{roninrektii}), and bugs introduced in new code (LI.FI~\cite{li_fi_incident_2023}). (3) \emph{poor security hygiene:} retained privileges by former developers (Infini~\cite{yona2025rogue}, Holograph~\cite{blockbasis2024}) and compromised private keys (OKX DEX~\cite{rekt2023okxdex}, Rho Market~\cite{olympix2024rho}) led to significant losses. (4) \emph{design oversights:} errors like the one in Yearn Finance~\cite{certik2023yearn} (incorrect address) or fundamental design flaws in logic or incentive mechanisms (Euler Finance’s liquidation discounts~\cite{slowmist2023euler}) were exploited. (5) \emph{ignoring warnings/known issues:} reusing code with known vulnerabilities or explicit warnings, as seen with EraLend~\cite{rekt2023eralend}, highlights a disregard for best practices.
\end{InsightBox}

\begin{InsightBox}[insight:cross-chain]{Complex Cross-Chain Exploits}
Three of the five largest losses (Nomad~\cite{nomad_rekt_2022}, Ronin~\cite{roninbridgeanalysis}, Rho Market~\cite{rekt2024rho}) exploited bridge or multi-chain logic, where state synchronicity is difficult to reason about formally. Emergence of these vulnerabilities calls for monitoring and analysis of cross-chain logic~\cite{HighGuard}
\end{InsightBox}

\begin{figure}[!t]
\centering
\includegraphics[width=1\columnwidth]{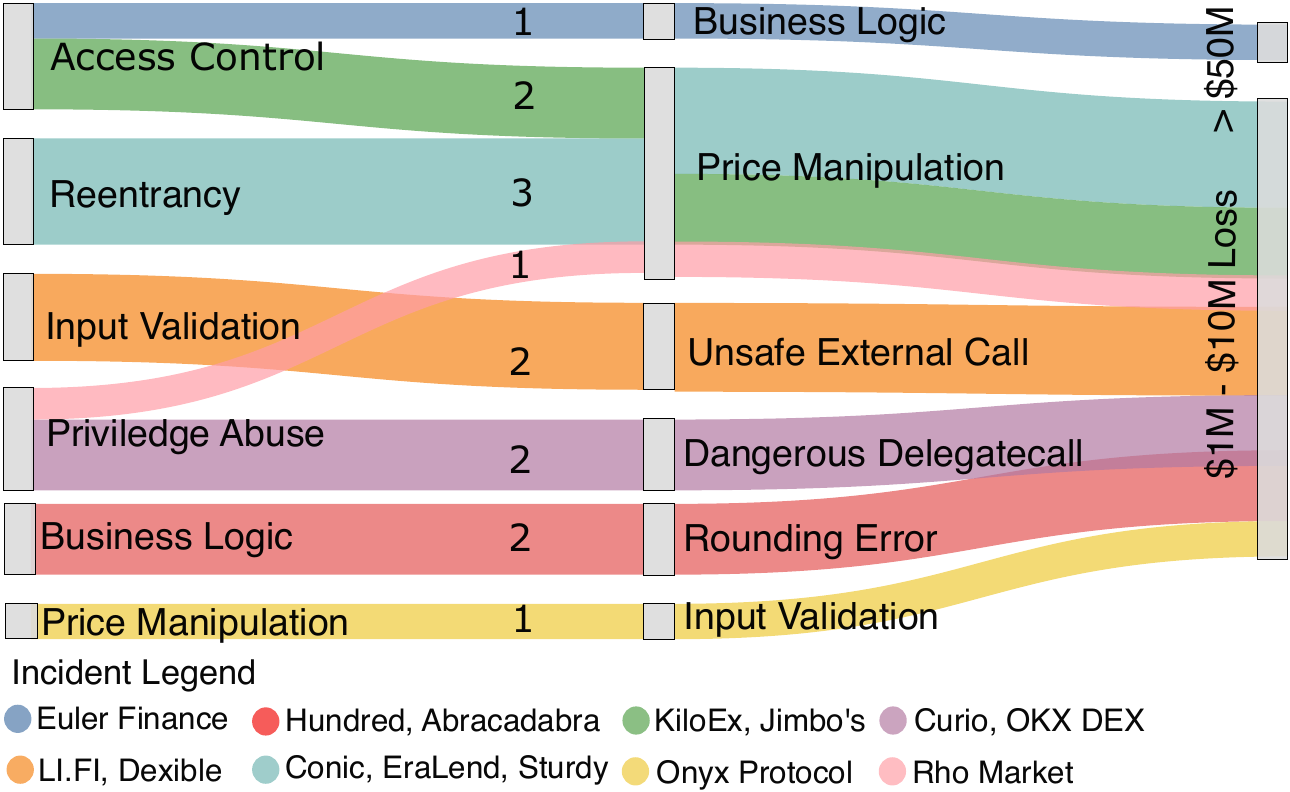}.
\caption{
\footnotesize
Incidents in which a chain of vulnerabilities was exploited. The first column (from left) represents the entry vulnerability used in an \emph{exploit chain}, and the second column shows the second vulnerability abused in the exploit chain of the incident (number on the flow presents the number of incidents involving this exploit chain flow).}
\label{fig:exploitChains}
\end{figure}

\begin{InsightBox}[insight:multi-vuln-incidents]{Multi-Vulnerability Exploit Chains}
\textcolor{blue!50!black}{\emph{Many incidents were not caused by an isolated flaw but rather a combination of weaknesses}} (e.g., rows \hyperlink{row1}{1}, \hyperlink{row18}{18}, \hyperlink{row19}{19}, \hyperlink{row47}{47}, \hyperlink{row48}{48} in \Cref{tab:main_attack_attributes}). For example, the Sturdy Finance attack~\cite{rekt2023sturdy,solidityscan2023sturdy} leveraged both read-only reentrancy and oracle manipulation to maximize extracted value, while the Jimbo protocol exploit~\cite{rekt_news_jimbo,numen_cyber_labs_jimbo} combined price manipulation with access control flaws.

Euler Finance involved both access control and business logic flaws~\cite{immunebytes2023euler, slowmist2023euler}, and LI.FI involved both access control and input validation issues~\cite{li_fi_incident_2023, solidityscan_li_fi_analysis_2023}. 
\Cref{fig:exploitChains} shows all of the incidents involving a chain of exploited vulnerabilities. As this figure shows, some vulnerabilities are only used to enable others to be exploited in an \textcolor{blue!50!black}{\emph{exploit chain}}. For instance, access control (\Cref{vuln:Access_Control}) in all exploit chains (3 cases) is performed to pull off the second step of the attack (price manipulation and business logic). Furthermore, \Cref{fig:exploitChains} shows the role of reentrancy in exploit chains: as the thickest flow in the diagram shows, it is most commonly used to create an entry during which the attacker executes the second step: price manipulation. According to these results, in the case of an exploit chain, price manipulation is always enabled with one of the following vulnerabilities in the contract: access control, reentrancy, and privilege abuse (human factor).
\end{InsightBox}

\begin{InsightBox}[insight:bad-emergency-responses]{Inadequate Emergency Response Mechanisms}
The Seneca protocol incident, where pause functionalities were declared \code{internal}-only, rendered them useless during the attack. Such incident response functionalities are rarely used in practice, which might result in their incompetence~\cite {rekt2024seneca}.
\end{InsightBox}

\begin{InsightBox}[insight:flash]{Flash Loans as Attack Amplifiers}
Flash loans were instrumental for price manipulation attacks (e.g., WooFi~\cite{olympix2023woofi}, Zunami Protocol~\cite{solidityscan2023zunami}), reentrancy (Fei Rari~\cite{rektnews2022}), and exploiting rounding errors (Sonne Finance~\cite{certik2024}). They provide attackers with significant temporary capital to execute exploits that would otherwise be impractical.
\end{InsightBox}

\begin{InsightBox}[insight:insider-threats]{Insider Threats and Privilege Abuse}
Access control issues were not just the result of external attackers but also about insufficient organizational controls and excessive privileges (Swaprum~\cite{cryptorank2023swaprum}).
\end{InsightBox}

\begin{InsightBox}[insight:oracle]{Oracle Security}
The high number of price manipulation incidents underscores that secure, manipulation-resistant oracles are crucial. Reliance on spot prices from single DEXs, insufficient liquidity in oracle pairs, or oracles whose prices can be influenced within a single atomic transaction remains a major weak point.
\end{InsightBox}

\begin{InsightBox}[insight:composability]{Composability Risks}
While DeFi's composability helps innovation, it also increases attack surfaces. Interactions between protocols (e.g., Abracadabra's cross-protocol state issue~\cite{halborn2025abracadabra}) or complex internal logic can hide subtle bugs. Composability-aware analysis and verification of protocols helps prevent these issues~\cite{ComposableVerification}.
\end{InsightBox}

\begin{InsightBox}[insight:new-evm-features]{New Attack Vectors for EVM Upgrades}
The SIR Trading incident~\cite{rekt2025sirtrading} which happened due to an attack vector on transient storage opcodes (EIP-1153~\cite{EIP1153}) demonstrates that new EVM features can introduce unforeseen vulnerabilities if not implemented and understood correctly by developers.
\end{InsightBox}

\begin{InsightBox}[insight:upgrdability]{Complications of Proxy/Upgradability Pattern}
Proxy or upgradability patterns allow for bug fixes but also introduce severe risks if admin keys are compromised (in case of OKX DEX~\cite{rekt2023okxdex} incident) or if governance mechanisms controlling upgrades are subverted (Tornado Cash~\cite{rekt2023tornado} and Atlantis Loans~\cite{solidityscan2023atlantis} incidents).
\end{InsightBox}

For a detailed review of each incident, refer to our accompanying repository\footnote{\scriptsize\href{https://github.com/HadisRe/SoK-Root-Cause-of-Smart-Contract-Incidents}{https://github.com/HadisRe/SoK-Root-Cause-of-Smart-Contract-Incidents}} providing in-depth analysis.

\section{A Multi-Tier Root-Cause Framework for Smart Contract Exploits}\label{sec:rq3-results}
We started by identifying smart contract vulnerability classes from academic literature (in \Cref{sec:rq1-results}) to understanding which classes of these vulnerabilities are indeed exploited in real-world attacks (in \Cref{sec:rq2-results}). 
Our analysis of these high-impact incidents revealed that financially severe incidents often stem from sources beyond simple developer mistakes. These exploits frequently originate from flawed economic designs, broken operational procedures (e.g., governance protocols), and incorrect assumptions about a contract's environment. Based on the observations from the real-world incident analyses and our comprehensive SLR of vulnerability classifications, we propose a novel, \emph{four}-tier root-cause-based framework.

The four-tier (not to be confused with blockchain layers) consist of: (1) \emph{Flawed Economic Design \& Protocol Logic}, where the core concept is unsound; (2) \emph{Protocol Lifecycle \& Governance Failures}, where human processes like upgrades, key management, or governance are broken; (3) \emph{External Dependency Vulnerabilities}, where the protocol makes unsafe assumptions about external data or contracts; and (4) \emph{Implementation-Level Weakness}, which encompasses classic coding errors. This model reveals that the most significant financial losses often stem from failures in the \textcolor{blue!50!black}{\emph{first two levels}}, \textcolor{blue!50!black}{\emph{areas that code-centric audits may overlook}}. This entails that in case of an existing flaw in an early tier (tier 1 and 2), the protocols will be exploitable regardless of quality of coding practices (lack of tiers 3 and 4 attack vectors). 
This model organizes exploit origins from the most fundamental (the design of the protocol) to the most specific (the implementation).
It recognizes that a contract can be technically sound yet economically flawed, or developed securely from known smart contract vulnerabilities but managed insecurely.
\Cref{tab:root_cause_framework} summarizes this framework, linking each category to the real-world incidents from our dataset of \num{50} attacks that exemplify it.

\begin{table*}[!t]
\caption{A multi-tier root-cause framework derived from real-world incidents. Incident IDs refer to rows in \Cref{tab:main_attack_attributes}.}
\label{tab:root_cause_framework}
\centering
\renewcommand{\arraystretch}{1.05} 
\setlength{\tabcolsep}{3pt} 
\scriptsize
\begin{tabular}{p{1.4cm} p{3.5cm} p{12.3cm}} 
\toprule
\textbf{Tier} & \textbf{Sub-Category} & \textbf{Description $\leftarrow$ Incidents Involved (ID)} \\
\midrule
\multirow{3}{=}{\textbf{T1:} Flawed Economic Design \& Protocol Logic} & 
Incorrect Pricing/Valuation Models & 
The protocol's core formulas for calculating value are inherently manipulable. \textit{Incidents: \hyperlink{row4}{4}, \hyperlink{row7}{7}, \hyperlink{row19}{19}, \hyperlink{row24}{24}, \hyperlink{row30}{30}, \hyperlink{row44}{44}, \hyperlink{row48}{48}} \\
\cmidrule{2-3}
& State Inconsistency \& Desynchronization & 
The logic allows different parts of the protocol to hold contradictory state information, enabling exploits. \textit{Incidents: \hyperlink{row20}{20}, \hyperlink{row28}{28}, \hyperlink{row45}{45}, \hyperlink{row47}{47}} \\
\cmidrule{2-3}
& Perverse Incentive Mechanisms & 
The design unintentionally rewards behavior that harms the protocol, such as self-liquidation. \textit{Incidents: \hyperlink{row1}{1}} \\
\midrule

\multirow{3}{=}{\textbf{T2:} Protocol Lifecycle \& Governance Failures} & 
Clumsy Deployments \& Upgrades & 
Mistakes during contract initialization or upgrades create vulnerabilities. \textit{Incidents: \hyperlink{row2}{2}, \hyperlink{row14}{14}, \hyperlink{row16}{16}, \hyperlink{row41}{41}} \\
\cmidrule{2-3}
& Compromised Keys \& Insider Threats & 
Administrative keys are stolen or abused by internal actors (current or former) to bypass security. \textit{Incidents: \hyperlink{row6}{6}, \hyperlink{row11}{11}, \hyperlink{row37}{37}, \hyperlink{row43}{43}, \hyperlink{row49}{49}} \\
\cmidrule{2-3}
& Governance System Exploitation & 
The rules of the DAO are manipulated to pass malicious proposals or seize administrative control. \textit{Incidents: \hyperlink{row34}{34}, \hyperlink{row40}{40}, \hyperlink{row46}{46}} \\
\midrule

\multirow{2}{=}{\textbf{T3:} External Dependency Vulnerabilities} & 
Oracle \& Price Feed Manipulation & 
The protocol uncritically trusts an external price feed that is susceptible to manipulation. \textit{Incidents: \hyperlink{row3}{3}, \hyperlink{row10}{10}, \hyperlink{row21}{21}, \hyperlink{row38}{38}} \\
\cmidrule{2-3}
& Unvalidated User/Contract Inputs & 
The protocol fails to sanitize or validate data and external calls originating from untrusted users. \textit{Incidents: \hyperlink{row13}{13}, \hyperlink{row15}{15}, \hyperlink{row18}{18}, \hyperlink{row29}{29}, \hyperlink{row31}{31}, \hyperlink{row32}{32}, \hyperlink{row35}{35}, \hyperlink{row42}{42}} \\
\midrule

\multirow{4}{=}{\textbf{T4:} Impl.-Level Vulnerabilities} & 
Interaction Pattern Failures & 
The code violates secure interaction patterns (e.g., CEI), enabling reentrancy. \textit{Incidents: \hyperlink{row5}{5}, \hyperlink{row8}{8}, \hyperlink{row25}{25}, \hyperlink{row33}{33}, \hyperlink{row50}{50}} \\
\cmidrule{2-3}
& Arithmetic Errors & 
The code is susceptible to over/underflow, or precision loss due to EVM integer math. \textit{Incidents: \hyperlink{row9}{9}, \hyperlink{row17}{17}, \hyperlink{row22}{22}, \hyperlink{row36}{36}, \hyperlink{row39}{39}} \\
\cmidrule{2-3}
& Flawed Access Control Logic & 
A flaw in the access control implementation (e.g., reversed parameters) allows it to be bypassed. \textit{Incidents: 
\hyperlink{row12}{12}, \hyperlink{row23}{23}, \hyperlink{row26}{26}} \\
\cmidrule{2-3}
& State \& Data Handling Errors & 
The code incorrectly manages data representation or state, leading to issues like storage collisions. \textit{Incidents: \hyperlink{row27}{27}} \\
\bottomrule
\end{tabular}
\end{table*}

\begin{table*}[htbp]
  \centering\scriptsize
  \caption{Defenses that neutralise four incidents in ``clumsy deployments \& upgrades'' sub-category.}
  \label{tab:l2-defenses}
  \begin{tabular}{@{}p{2.8cm}p{9.8cm}p{4.7cm}@{}}
    \toprule
    \textbf{Defense} & \textbf{Example} & \textbf{Effect on Incidents (\hyperlink{2}{2}/\hyperlink{14}{14}/\hyperlink{16}{16}/\hyperlink{41}{41})} \\
    \midrule
    Invariant‑checked upgrades & Foundry\cite{Foundry}/Echidna\cite{Echidna} test asserting post‑upgrade state ($\neq 0$, correct addresses, guard present). & Blocks deployment; test fails on first run. \\
    Staged roll‑outs & Deploy first on canary fork $\rightarrow$ 24h observation $\rightarrow$ mainnet. & Limits loss to canary TVL. \\
    Operational gates & Multi‑sig + two‑step \textit{pause $\rightarrow$ upgrade $\rightarrow$ un‑pause}. & Human review spots violating states. \\
    Runtime monitors & Circuit‑breaker on abnormal withdrawals / mint rates / external calls. & Halts draining within the first block. \\
    \bottomrule
  \end{tabular}
\end{table*}

\subsection{Tier 1: Flawed Economic Design \& Protocol Logic}
Here, the protocol is vulnerable not because of a bug, but because its core concept is economically or logically unsound (\num{12} incidents trace back to these vulnerabilities). The code may execute exactly as intended, but the intended functionality itself is flawed. For example, the academic vulnerability of \emph{business logic flaws} (where the contract does not match expected behavior) (\Cref{vuln:Business_logic_errors}) is a subset of this tier. However, our findings show this category is broader than business logic flaws, including the flawed liquidation incentive model in the {Euler Finance}~\cite{rekt2023euler} case and the insecure pricing formula in the {WooFi}~\cite{rekt_woofi_2024} exploit, which are design-level errors, not simple implementation bugs.

\subsection{Tier 2: Protocol Lifecycle \& Governance Failures}\label{sec:layer-2}
Our empirical data reveals that a significant number of catastrophic incidents (\num{12} incidents) originate from failures in the human and operational processes surrounding a smart contract. This tier is almost entirely absent from traditional, code-centric vulnerability taxonomies, yet it is a dominant cause of failure. Incidents like the {Nomad Bridge} exploit, caused by a clumsy upgrade procedure~\cite{nomad_rekt_2022}, or the {Infini} exploit, resulting from a failure to revoke developer privileges~\cite{yona2025rogue}, would not have been detected by existing code analysis tools. They represent failures in deployment, maintenance, and governance procedures that show the need \textcolor{blue!50!black}{\emph{to audit not just the code, but also the team's operational security}}.

\subsection{Tier 3: External Dependency Vulnerabilities}
This tier addresses vulnerabilities and incidents arising from a contract's data dependencies with its environment (12 incidents). The contract's internal logic may be sound, but it makes unsafe assumptions about the inputs it receives from the outside world. This expands upon the literature's concept of \emph{Untrustworthy Data Feeds} (\Cref{vuln:Untrustworthy_Data_Feeds}) to include any external dependency. This includes not only price oracles~\cite{OracleExploitation}, as seen in the {Inverse Finance} attack~\cite{rekt2022inversefinance}, but also unvalidated \code{calldata} from users, as exploited in the {LI.FI} incident~\cite{li_fi_incident_2023}. Furthermore, our analysis identifies the critical role of environmental factors like the availability of \textcolor{blue!50!black}{\emph{flash loans, which, while not a vulnerability themselves, act as powerful amplifiers for exploits in this tier}}.

\subsection{Tier 4: Implementation-Level Weakness}
This tier aligns most closely with the traditional vulnerability classifications identified in our SLR (RQ1 results in \Cref{sec:rq1-results}). We traced back the origin of \num{14} incidents to these vulnerabilities (\Cref{tab:root_cause_framework}). 
These are concrete programming mistakes where the contract contains a known vulnerability. This includes well-documented issues such as {reentrancy} (\Cref{vuln:Reentrancy}), as seen in many attacks such as {Fei Rari} exploit~\cite{fei2023msmart}, and {integer over/underflow} (\Cref{vuln:integer-overflow}), found in multiple incidents such as {Velocore}~\cite{rekt2024velocore}. While critically important, the number of incidents that their root-cause is attributed to other tiers (see \Cref{tab:root_cause_framework}) suggests an incomplete picture of real-world attacks solely based on tier 4 vulnerabilities.

\subsection{A Case Study for Tier 2}
\label{sec:case-study}

In order to investigate the usefulness of the framework proposed, we analyze ``clumsy deployments and upgrades''--a sub-category of tier 2 vulnerabilities (from \Cref{tab:root_cause_framework}). This sub-category consists of contract upgrades and initialization that lead to vulnerabilities. 
Four of the largest bridge, vault, and DEX catastrophes in our dataset
(IDs \hyperlink{row2}{2}, \hyperlink{row14}{14}, \hyperlink{row16}{16}, \hyperlink{row41}{41}; \$\num{215.5}M combined loss) share the \emph{same}
root‑cause: a \emph{seemingly} flawless deployment and upgrade that left
the contract in an unsafe state while the byte‑code itself remained
bug‑free.  Treating them jointly shows that our sub‑category is the
\emph{minimal sufficient grain} at which attacker assumptions,
exploit steps, and defenses can be specified once and reused to prevent future exploits to similar protocols.

\subsubsection{Attacker Model}
\label{sec:l2-attacker-model}

Based on our incident analyses, we envision an attacker with the following attributes: 

\noindent\textbf{Goal:} Extract maximum value by exploiting mis‑configured contract state after upgrade or deployment. 

\noindent\textbf{Capabilities:} i) Craft arbitrary on‑chain transactions (privileged keys not required). ii) Inspect contract storage. iii) Use flash loans to amplify the effect. iv) Reads the verified source code of the protocol/ABI. 

\noindent\textbf{Aussumption:} Upgrades already live as contracts on mainnet.

\subsubsection{Exploit Walk‑Through}
\label{sec:l2-exploit} 
The incidents can be broken down into three steps: 

\noindent\textbf{Step A (State discovery):} Attacker reads mis‑initialised
variable (Nomad \code{committedRoot}, Ronin
\code{\_totalOperatorWeight}, Yearn vault basket, Unizen call‑guard)
and confirms it lifts a safety check.  

\noindent\textbf{Step B (Trigger):} Crafts 
transaction exploiting the
lifted check: replay any root (Nomad), sign with tiny weight (Ronin),
mint vault shares (Yearn), or route trade through the attacker contract
(Unizen).  

\noindent\textbf{Step C (Extraction):} Drains bridge funds / swaps inflated
shares / spends user approvals.  No reentrancy or oracle tricks are
needed; the upgrade itself is the point of failure.

\subsubsection{Defense Model}
\label{sec:l2-defenses} \Cref{tab:l2-defenses} presents our proposed defenses against the attacker model of this sub-category. 
Before moving real assets, effective tier 2 security
must wrap the \emph{entire upgrade life‑cycle}: (i) \emph{pre‑deployment/upgrade} initiation
property‑based tests that abort if critical state is mis‑initialised;
(ii) a \emph{canary rollout} that risks only a capped pilot TVL while the new code faces live mainnet conditions for \num{24}h; (iii) human \emph{multi‑sig gating} that forces a pause‑upgrade‑unpause ritual; and (iv) on‑chain \emph{circuit‑breakers}~\cite{captureDCR} that freeze abnormal withdrawals, mint rates, or delegate‑calls within the first block.

\section{Conclusion and Future Work}
\label{sec:conclusion}

\paragraph{Threats to Validity} Our evidence base contains only incidents that were 
\emph{publicly revealed} and \emph{financially substantial}, so \emph{silent failures} and \emph{minor incidents} remained outside the scope of this work which can affect our results. All data came from the EVM blockchain family, which could lead to an alternative execution platform exposing additional incident patterns not observed in our study. 

\paragraph{Future Directions} At the economic layer, the community needs agent‑based simulation, model‑checking, and other formal‑methods machinery that can prove the designed system preserves economic soundness. At the governance and lifecycle tier, development pipelines should integrate upgrade simulators, proposal‑linting tools and role‑minimisation checks. External data dependency analysis can evolve into \emph{whole‑mempool} and \emph{cross‑protocol reasoning} that surfaces oracle manipulation or flash‑loan amplification before execution. Running across all tiers, real‑time monitoring observing mempool based on exploit‑chain-aware rules should be investigated.
Finally, as bridges and roll‑ups are common among loss statistics (\Cref{tab:main_attack_attributes}), they require more investigation as the contracts interacting are heterogeneous~\cite{HighGuard}.

\paragraph{Conclusion} This study unites the literature's focus on implementation bugs with the practitioner's day‑to‑day experience of successful attacks on Ethereum‑based protocols.  Through a systematic literature review of \papersNum{} peer‑reviewed papers and dissecting \num{50} publicly disclosed exploits that caused a total loss of \${\totalLosses} between 2022 and 2025, we show that less than one‑third of recorded losses can be traced to an isolated coding mistake. The remainder cascaded from flawed incentive design, clumsy governance decisions, unsafe assumptions about external dependencies, or exploit chains-factors that mainstream static and dynamic analyzers rarely take into account. To capture this, we introduced a four‑tier root‑cause framework ranging from economic logic at tier 1 to programming mistakes at tier 4; together with our labeled incident dataset and mitigation catalog, the framework equips auditors, tool builders and protocol designers with a map of the modern threat surface based on real incidents.

\appendix

\setcounter{table}{0}
\renewcommand{\thetable}{A.\arabic{table}}

\begin{table*}[!t]
\centering
\scriptsize
\caption{Mapping of exploit transaction hashes, compromised victim contracts, and attacker addresses (smart contracts and EOAs).}
\label{tab:attack_addresses}
\begin{tabular}{@{} c p{0.33\textwidth} p{0.33\textwidth} p{0.22\textwidth} @{}}
\toprule
\textbf{ID} & \textbf{Attack Transactions} & \textbf{Victim Contract} & \textbf{Attacker Smart Contract/EOA} \\
\midrule

1 & \href{https://etherscan.io/tx/0xc310a0affe2169d1f6feec1c63dbc7f7c62a887fa48795d327d4d2da2d6b111d}{0xc310}, \href{https://etherscan.io/tx/0x62bd3d31a7b75c098ccf28bc4d4af8c4a191b4b9e451fab4232258079e8b18c4}{0x62bd}, \href{https://etherscan.io/tx/0x3097830e9921e4063d334acb82f6a79374f76f0b1a8f857e89b89bc58df1f311}{0x3097}, \href{https://etherscan.io/tx/0x47ac3527d02e6b9631c77fad1cdee7bfa77a8a7bfd4880dccbda5146ace4088f}{0x47ac} & \href{https://etherscan.io/address/0xe025e3ca2be02316033184551d4d3aa22024d9dc}{0xe025} &   \href{https://etherscan.io/address/0xeBC29199C817Dc47BA12E3F86102564D640CBf99}{0xebc2}, \href{https://etherscan.io/address/0x036cec1a199234fc02f72d29e596a09440825f1c}{0x036c}, \href{https://etherscan.io/address/0xD3b7CEA28Feb5E537fcA4E657e3f60129456eaF3}{0xd3b7}, \href{https://etherscan.io/address/0x0b812c74729b6aBc723F22986C61D95344ff7ABA}{0x0b81}\\

\hline
2 & \href{https://etherscan.io/tx/0xa5fe9d044e4f3e5aa5bc4c0709333cd2190cba0f4e7f16bcf73f49f83e4a5460}{0xa5fe} & \href{https://etherscan.io/address/0x88a69b4e698a4b090df6cf5bd7b2d47325ad30a3}{0x88a6} & \href{https://etherscan.io/address/0xb5c55f76f90cc528b2609109ca14d8d84593590e}{0xb5c5} \\
\hline
3 & \href{https://polygonscan.com/tx/0x31957ecc43774d19f54d9968e95c69c882468b46860f921668f2c55fadd51b19}{0x3195}
  & \href{https://polygonscan.com/address/0x8f55d884cad66b79e1a131f6bcb0e66f4fd84d5b}{0x8f55} & \href{https://polygonscan.com/address/0xcacf2d28b2a5309e099f0c6e8c60ec3ddf656642}{0xcacf} \\

 \hline
4  &\href{https://arbiscan.io/tx/0xe80a16678b5008d5be1484ec6e9e77dc6307632030553405863ffb38c1f94266}{0xe80a}
 & \href{https://arbiscan.io/address/0xeff23b4be1091b53205e35f3afcd9c7182bf3062}{0xeff2} & \href{https://arbiscan.io/address/0x9961190b258897bca7a12b8f37f415e689d281c4}{0x9961}  \\
 \hline
5 & \href{https://etherscan.io/tx/0xd9ee4fc5ee0b8815e6aae20e8bc5697ee49b8a1c76619a008bf534a4084197dc}{0xd9ee}, \href{https://etherscan.io/tx/0xadbe5cf9269a001d50990d0c29075b402bcc3a0b0f3258821881621b787b35c6}{0xadbe}, \href{https://etherscan.io/tx/0x0f75349606610313cb666277eeda612e72be624cae061d017e503056bbf4d8e0}{0x0f75}, \href{https://etherscan.io/tx/0x0742b138a78ad9bd5d0b55221d514637313bc64c40272ca98c8d0417a519e2e4}{0x0742}, \href{https://etherscan.io/tx/0x254735c6c14e4d338b1cc5bca43aab6b0f395ae06085013b1b2527180d270a31}{0x2547}, \href{https://etherscan.io/tx/0xab486012f21be741c9e674ffda227e30518e8a1e37a5f1d58d0b0d41f6e76530}{0xab48} & 
\href{https://etherscan.io/address/0x32075bad9050d4767018084f0cb87b3182d36c45}{0x3207} & \href{https://etherscan.io/address/0xE39f3C40966DF56c69AA508D8AD459E77B8a2bc1}{0xE39f}, \href{https://etherscan.io/address/0x32075bad9050d4767018084f0cb87b3182d36c45}{0x3207} 
\\
\hline
6 & \href{https://etherscan.io/tx/0xa5fe9d044e4f3e5aa5bc4c0709333cd2190cba0f4e7f16bcf73f49f83e4a5460}{0xa5fe} & \href{https://etherscan.io/address/0x9A79f4105A4e1A050Ba0b42F25351D394fA7E1DC}{0x9a79} & \href{https://etherscan.io/address/0xc49b5e5b9da66b9126c1a62e9761e6b2147de3e1}{0xc49b} \\
\hline
7  & \href{https://etherscan.io/tx/0x485e08dc2b6a4b3aeadcb89c3d18a37666dc7d9424961a2091d6b3696792f0f3}{0x485e}, \href{https://etherscan.io/tx/0x09a3a12d58b0bb80e33e3fb8e282728551dc430c65d1e520fe0009ec519d75e8}{0x09a3}, \href{https://etherscan.io/tx/0x396a83df7361519416a6dc960d394e689dd0f158095cbc6a6c387640716f5475}{0x396a} & 
  \href{https://etherscan.io/address/0xaf2acf3d4ab78e4c702256d214a3189a874cdc13}{0xaf2a} & \href{https://etherscan.io/address/0x50275e0b7261559ce1644014d4b78d4aa63be836}{0x5027}, \href{https://etherscan.io/address/0xc9b826bad20872eb29f9b1d8af4befe8460b50c6}{0xc9b8}
  \\

\hline
8 & \href{https://etherscan.io/tx/0x7e7f9548f301d3dd863eac94e6190cb742ab6aa9d7730549ff743bf84cbd21d1}{0x7e7f}, \href{https://etherscan.io/tx/0xfda0dde38fa4c5b0e13c506782527a039d3a87f93f9208c104ee569a642172d2}{0xfda0}, \href{https://etherscan.io/tx/0x7961b0d3382bddff4c777e432902ea0b6940414ec16aa1a1dcebc4ebcbd8f867}{0x7961}, \href{https://etherscan.io/tx/0xca87f257280e19378dc1890a478514195f068857affacde0b92c851b897dff9e}{0xca87} & - &   \href{https://etherscan.io/address/0x7a2f4d625fb21f5e51562ce8dc2e722e12a61d1b}{0x7a2f}\\
\hline
9 & \href{https://optimistic.etherscan.io/tx/0x9312ae377d7ebdf3c7c3a86f80514878deb5df51aad38b6191d55db53e42b7f0}{0x9312} & \href{https://optimistic.etherscan.io/address/0xe3b81318b1b6776f0877c3770afddff97b9f5fe5}{sovelo}, \href{https://optimistic.etherscan.io/address/0xec8fea79026ffed168ccf5c627c7f486d77b765f}{sousdc}, \href{https://optimistic.etherscan.io/address/0xf7b5965f5c117eb1b5450187c9dcfccc3c317e8e}{soweth} & \href{https://optimistic.etherscan.io/address/0x02fa2625825917e9b1f8346a465de1bbc150c5b9}{0x02fa}  \\
\hline
 10 & \href{https://etherscan.io/tx/0x561e94c8040c82f8ec717a03e49923385ff6c9e11da641fbc518ac318e588984}{0x561e}, \href{https://etherscan.io/tx/0x20a6dcff06a791a7f8be9f423053ce8caee3f9eecc31df32445fc98d4ccd8365}{0x20a6}, \href{https://etherscan.io/tx/0x600373f67521324c8068cfd025f121a0843d57ec813411661b07edc5ff781842}{0x6003} 
& \href{https://etherscan.io/address/0x39b1dF026010b5aEA781f90542EE19E900F2Db15}{0x39b1}  & \href{https://etherscan.io/address/0x117C0391B3483E32AA665b5ecb2Cc539669EA7E9}{0x117c}, \href{https://etherscan.io/address/0x8B4C1083cd6Aef062298E1Fa900df9832c8351b3}{0x8b4c}\\
\hline
11 & \href{https://etherscan.io/tx/0x0cc143ccf3316d47b36a2e45577922f4ebe2374966bb22c1e9cf49c747d46396}{0x0cc1} &- &- \\
\hline
 12 & \href{https://etherscan.io/tx/0x6129dd42778345bc278822a7feadeacb933f5e56ce51114e686832ad239307a8}{0x6129} & \href{https://arbiscan.io/address/0x189cf534de3097c08b6beaf6eb2b9179dab122d1}{0xbc1b} & \href{https://arbiscan.io/address/0xe2ee6252509382a2b6504d5a5f7a1c5018a38168}{0xe2ee}  \\
 \hline
13 & \href{https://explorer.mode.network/tx/0x9aa3fd43a6b0f85b4f1bf74f0c9e79773f238591d9c6fe666287bd2c8ac19009}{0x9aa3}, \href{https://explorer.mode.network/tx/0x37e53b15cb7f298bd8c45fcbbd914ba90feb3946f5511fc55bc986b7472956df}{0x37e5}, \href{https://explorer.mode.network/tx/0x5db6d90a17a44bed6d9ed9ca73d800df2661751fa1a273e71fc2174ad3b6944f}{0x5db6} & \href{https://explorer.mode.network/address/0x964dd444e3192f636322229080a576077b06fba3}{0x964d} &   \href{https://explorer.mode.network/address/0x9E34d89C013Da3BF65fc02b59B6F27D710850430}{0x9e34} \\
\hline
14 & \href{https://etherscan.io/tx/0x2619570088683e6cc3a38d93c3d98899e5783864e15525d5f5810c11189ba6cb}{0x2619} & \href{https://etherscan.io/address/0xfc274ec92bbb1a1472884558d1b5caac6f8220ee}{0xfc27} & - \\
\hline
15 & \href{https://etherscan.io/tx/0xd82fe84e63b1aa52e1ce540582ee0895ba4a71ec5e7a632a3faa1aff3e763873}{0xd82f}, \href{https://etherscan.io/tx/0x65a92b189e4ae0b8a8a02cd59c5e9f6832586bd5167d41a24eb4f4d2ac692755}{0x65a9}   & \href{https://etherscan.io/address/0xf28a352377663ca134bd27b582b1a9a4dad7e534#code}{0xf28a} & \href{https://etherscan.io/address/0x7742ed59e9ecf1712bc4c6bdd0c526e903a7f2c8}{0x7742}  \\
\hline
16 & \href{https://etherscan.io/tx/0xd55e43c1602b28d4fd4667ee445d570c8f298f5401cf04e62ec329759ecda95d}{0xd55e}, \href{https://etherscan.io/tx/0x8db0ef33024c47200d47d8e97b0fcfc4b51de1820dfb4e911f0e3fb0a4053138}{0x8db0} & \href{ }{-} & \href{https://etherscan.io/address/0x5bac20beef31d0eccb369a33514831ed8e9cdfe0}{0x5bac}, \href{https://etherscan.io/address/0x16af29b7efbf019ef30aae9023a5140c012374a5}{0x16af}  \\
\hline
17 & \href{https://starkscan.co/tx/0x0160a5841b3e99679691294d1f18904c557b28f7d5fe61577e75c8931f34a16f}{0x0160} & \href{https://starkscan.co/contract/0x04c0a5193d58f74fbace4b74dcf65481e734ed1714121bdc571da345540efa05}{0x04c0} & \href{https://starkscan.co/contract/0x04d7191dc8eac499bac710dd368706e3ce76c9945da52535de770d06ce7d3b26}{0x04d7} \\
\hline
18 & \href{https://etherscan.io/tx/0xd82fe84e63b1aa52e1ce540582ee0895ba4a71ec5e7a632a3faa1aff3e763873}{0xd82f}, \href{https://etherscan.io/tx/0x65a92b189e4ae0b8a8a02cd59c5e9f6832586bd5167d41a24eb4f4d2ac692755}{0x65a9} & \href{https://etherscan.io/address/0xc74fc202a6d0cf7a0ac38e99607b8629cdf2cb10}{0xc74f}& \href{https://etherscan.io/address/0x7742ed59e9ecf1712bc4c6bdd0c526e903a7f2c8}{0x7742}  \\

\hline
19 & \href{https://optimistic.etherscan.io/tx/0x6e9ebcdebbabda04fa9f2e3bc21ea8b2e4fb4bf4f4670cb8483e2f0b2604f451}{0x6e9e}, \href{https://optimistic.etherscan.io/tx/0x15096dc6a59cff26e0bd22eaf7e3a60125dcec687580383488b7b5dd2aceea93}{0x1509}  & \href{https://optimistic.etherscan.io/address/0x8ADB8131C5F951C61C90418885777DCC9A00728D}{0x8adb} & \href{https://optimistic.etherscan.io/address/0x155da45d374a286d383839b1ef27567a15e67528}{0x155d}\\
\hline
20 & \href{https://etherscan.io/tx/0x26a83db7e28838dd9fee6fb7314ae58dcc6aee9a20bf224c386ff5e80f7e4cf2}{0x26a8}, \href{https://etherscan.io/tx/0xdb4616b89ad82062787a4e924d520639791302476484b9a6eca5126f79b6d877}{0xdb46} & \href{https://etherscan.io/address/0x7259e152103756e1616A77Ae982353c3751A6a90}{0x7259} & \href{https://etherscan.io/address/0x87f585809ce79ae39a5fa0c7c96d0d159eb678c9}{0x87f5} \\
\hline

21 & \href{https://arbiscan.io/tx/0xc523c6307b025ebd9aef155ba792d1ba18d5d83f97c7a846f267d3d9a3004e8c}{0xc523} & \href{https://arbiscan.io/address/0x7596ACadf6c93f01b877F5A44b49407ffFC53508}{0x7596} & \href{https://arbiscan.io/address/0xc29d94386ff784006ff8461c170d1953cc9e2b5c}{0xc29d} \\
\hline
22 & \href{https://etherscan.io/tx/0x04975648e0db631b0620759ca934861830472678dae82b4bed493f1e1e3ed03a}{0x0497}, \href{https://etherscan.io/tx/0xb5c94efa0c8fd8f5c8cc2826e374a99620b01061d395b59b8f45dddc9fce1c60}{0xb5c9}, \href{https://etherscan.io/tx/0xb16bbf03d324b66685c94d62dbe31c739ee23c114b3915d169c74cd7c98eec8c}{0xb16b}, \href{https://etherscan.io/tx/0x3947e5a4d98104e313e08ee321673e1183db3d6ff8b7207f3eabb36f71436c1d}{0x3947}, \href{https://etherscan.io/tx/0x9ce5187c7160f531189e4765f21af5975dc2a62d961fb61ae09866d082918256}{0x9ce5}, \href{https://etherscan.io/tx/0xb0688eb1f46c28f36d7397366146fced23d3f8da7e08b760a5f612ce134ee9d2}{0xb068}, \href{https://etherscan.io/tx/0x62734ce80311e64630a009dd101a967ea0a9c012fabbfce8eac90f0f4ca090d6}{0x6273} & \href{https://etherscan.io/address/0xb02f39e382c90160eb816de5e0e428ac771d77b5}{0xb02f} & \href{https://etherscan.io/address/0x019bfc71d43c3492926d4a9a6c781f36706970c9}{0x019b} \\

\hline
23 & \href{https://etherscan.io/tx/0x39328ea4377a8887d3f6ce91b2f4c6b19a851e2fc5163e2f83bbc2fc136d0c71}{0x3932} & \href{https://etherscan.io/address/0xa35f69899796ddbc4a8904511d2f1f040b779cb7}{0xa35f} & \href{https://etherscan.io/address/0xA3a64255484aD65158AF0F9d96B5577F79901a1D}{0xa3a6} \\

\hline
24 & \href{https://arbiscan.io/tx/0x025cf2858723369d606ee3abbc4ec01eab064a97cc9ec578bf91c6908679be75}{0x025c} & \href{https://arbiscan.io/address/0x4b57adc00ac38f74506d29fc4080e3dc65b78a69}{0x4b57} & \href{https://arbiscan.io/address/0x5351536145610aa448a8bf85ba97c71caf31909c}{0x5351} \\
\hline
25 & \href{https://arbiscan.io/tx/0x025cf2858723369d606ee3abbc4ec01eab064a97cc9ec578bf91c6908679be75}{0x8b74} & \href{https://etherscan.io/address/0x8d67db0b205e32a5dd96145f022fa18aae7dc8aa}{0x8d67} & \href{https://etherscan.io/address/0x10db234e02c3889d8e408c7084e8ce10892bdad7}{0x10db} \\
\hline
26 & \href{https://basescan.org/tx/0x6b378c84aa57097fb5845f285476e33d6832b8090d36d02fe0e1aed909228edd}{0x6b37}  & \href{https://etherscan.io/address/0x551f3110f12c763D1611d5A63B5F015d1c1a954C}{0x551f} & \href{https://basescan.org/address/0x00fac92881556a90fdb19eae9f23640b95b4bcbd}{0x00fa}\\
\hline
27 & \href{https://etherscan.io/tx/0xa05f047ddfdad9126624c4496b5d4a59f961ee7c091e7b4e38cee86f1335736f}{0xa05f} & \href{https://etherscan.io/address/0xb91ae2c8365fd45030aba84a4666c4db074e53e7}{0xb91a} & \href{https://etherscan.io/address/0x27defcfa6498f957918f407ed8a58eba2884768c}{0x27de} \\
\hline
28 & \href{https://arbiscan.io/tx/0xed17089aa6c57b7d5461209e853bdb56bc3460a91805e20d2590609a515ef0b0}{0xed17} & \href{https://arbiscan.io/address/0x625Fe79547828b1B54467E5Ed822a9A8a074bD61}{0x625F} & \href{https://arbiscan.io/address/0xAF9e33Aa03CAaa613c3Ba4221f7EA3eE2AC38649}{0xaf9e} \\
\hline
29 & \href{https://arbiscan.io/tx/0x9efe855cd3783462207ff8a3d94dc17a74e2b2f00bf1b4c8a7e0135dae83ab5c}{0x9efe}  & \href{https://arbiscan.io/address/0x52ee5c0ea2e7b38d4b24c09d4d18cba6c293200e}{0x52ee} & \href{https://arbiscan.io/address/0xb87881637b5c8e6885c51ab7d895e53fa7d7c567}{0xb878}\\
\hline
30 & \href{https://etherscan.io/tx/0x46567c731c4f4f7e27c4ce591f0aebdeb2d9ae1038237a0134de7b13e63d8729}{0x4656}  & \href{ }{-} & \href{https://etherscan.io/address/0x680910cf5fc9969a25fd57e7896a14ff1e55f36b}{0x6809}\\
\hline

31 & \href{https://etherscan.io/tx/0x138daa4cbeaa3db42eefcec26e234fc2c89a4aa17d6b1870fc460b2856fd11a6}{0x138d} & \href{https://etherscan.io/address/0xde62e1b0edaa55aac5ffbe21984d321706418024#code}{0xde62} & \href{https://etherscan.io/address/0x684083f312ac50f538cc4b634d85a2feafaab77a}{0x6840} \\
\hline
 32 & \href{https://etherscan.io/tx/0x92cdcc732eebf47200ea56123716e337f6ef7d5ad714a2295794fdc6031ebb2e}{0x92cd} & \href{https://etherscan.io/address/0x534a3bb1ecb886ce9e7632e33d97bf22f838d085}{0x534a}, \href{https://etherscan.io/address/0x53d2D9D2b36d2784D80297E6532e3BD965435021}{0x53d2}, \href{https://etherscan.io/address/0x0d2b0c59D6a51eAE239A6C6eE29cFE73b79cC35b}{0x0d2b}, \href{https://etherscan.io/address/0x8f690502964348acbab0E3E3E81192A582715d89}{0x8f69}, \href{https://etherscan.io/address/0xC581d6Ef24146f745d5Bc014cc114C8F0CA74783}{0xc581}, \href{https://etherscan.io/address/0x460fDebe3D0B26e9DC194De80B3111b369B29272}{0x460f}, \href{https://etherscan.io/address/0x3e9F59f371dB249F2D95cf1E6F5224Ff1D7328Ab}{0x3e9f}, \href{https://etherscan.io/address/0x52F0b1B251FCa6f2b1191803a0aA4d9f4dB6F924}{0x52f0} & \href{https://etherscan.io/address/0x67104175fc5fabbdb5A1876c3914e04B94c71741}{0x6710} \\
 \hline
 33 & \href{https://basescan.org/tx/0x8fcdfcded45100437ff94801090355f2f689941dca75de9a702e01670f361c04}{0x8fcd} & \href{https://basescan.org/address/0x6a0b87d6b74f7d5c92722f6a11714dbeda9f3895#code}{0x6a0b} & \href{https://basescan.org/address/0x012Fc6377F1c5CCF6e29967Bce52e3629AaA6025}{0x012f} \\
\hline
34 & \href{https://etherscan.io/tx/0x3274b6090685b842aca80b304a4dcee0f61ef8b6afee10b7c7533c32fb75486d}{0x3274} & \href{https://etherscan.io/address/0xc503893b3e3c0c6b909222b45f2a3a259a52752d}{0xc503} & \href{https://etherscan.io/address/0x092123663804f8801b9b086b03b98d706f77bd59}{0x0921} , \href{https://etherscan.io/address/0x592340957ebc9e4afb0e9af221d06fdddf789de9}{0x5923}\\
\hline

35 & \href{https://etherscan.io/tx/0xb2e3ea72d353da43a2ac9a8f1670fd16463ab370e563b9b5b26119b2601277ce}{0xb2e3} & \href{https://etherscan.io/address/0xcff07c4e6aa9e2fec04daaf5f41d1b10f3adadf4}{0xcff0} & \href{https://etherscan.io/address/0xcff07c4e6aa9e2fec04daaf5f41d1b10f3adadf4}{0xcff0} \\
\hline
36 & \href{https://etherscan.io/tx/0xfeedbf51b4e2338e38171f6e19501327294ab1907ab44cfd2d7e7336c975ace7}{0xfeed}  & \href{https://etherscan.io/address/0x9ab6b21cdf116f611110b048987e58894786c244}{0x9ab6}, \href{https://etherscan.io/address/0xd0db31473caad65428ba301d2174390d11d0c788}{0xd0db}& \href{https://etherscan.io/address/0x0A3340129816a86b62b7eafD61427f743c315ef8}{0x0a33}, \href{https://etherscan.io/address/0xfdc0feaa3f0830aa2756d943c6d7d39f1d587110}{0xfdc0}, \href{https://etherscan.io/address/0x011992114806e2c3770df73fa0d19884215db85f}{0x0119} \\
\hline
 37 & \href{ }{-} & \href{https://scrollscan.com/address/0x2a9c973a2f5cb494eA84Fd0811aA7701f4d56401}{0x2a9c} & \href{https://scrollscan.com/address/0x893D5C3E6d84785a648902e31f8734d2be648CdC}{0x893d} \\
  \hline
  
38 & \href{https://etherscan.io/tx/0x242a0fb4fde9de0dc2fd42e8db743cbc197ffa2bf6a036ba0bba303df296408b}{0x242a0f}, \href{https://etherscan.io/tx/0xb3f067618ce54bc26a960b660cfc28f9ea0315e2e9a1a855ede1508eb4017376}{0xb3f0}, \href{https://etherscan.io/tx/0xca1bbf3b320662c89232006f1ec6624b56242850f07e0f1dadbe4f69ba0d6ac3}{0xca1b} & \href{}{-} & \href{https://etherscan.io/address/0x841ddf093f5188989fa1524e7b893de64b421f47}{0x841d} \\

\hline
39 & \href{https://lineascan.build/tx/0xed11d5b013bf3296b1507da38b7bcb97845dd037d33d3d1b0c5e763889cdbed1}{0xed11d}, \href{https://lineascan.build/tx/0x37434e674efc4e7cfeed7746095301ace5636028906fe548b786ead286e35eb0}{0x3743}, \href{https://explorer.zksync.io/tx/0x4156d73cadc18419220f5bcf10deb4d97a3d3f7533d63ba90daeabc5fd11ba17}{0x4156}
& \href{https://lineascan.build/address/0xed4e130f6f9e68918996f7e1e46a3306b3e12cec}{0xed4e} ,
\href{https://lineascan.build/address/0xb7f6354b2cfd3018b3261fbc63248a56a24ae91a}{0xb7f6} ,
\href{https://lineascan.build/address/0xc030fba4b741b770f03e715c3a27d02c41fc9dae}{0xc030}, 
\href{https://explorer.zksync.io/address/0xf7f76b30a301524fe76508546B1e3762eF2B9267}{0xf7f7} 
& \href{https://lineascan.build/address/0x8cdc37ed79c5ef116b9dc2a53cb86acaca3716bf}{0x8cdc} \\
\hline
40 & \href{https://etherscan.io/tx/0x4ff4028b03c3df468197358b99f5160e5709e7fce3884cc8ce818856d058e106}{0x4ff40} & \href{ }{-} & \href{https://etherscan.io/address/0xdaAa6294C47b5743BDafe0613d1926eE27ae8cf5}{0xdaaa} \\
\hline
41 & \href{https://etherscan.io/tx/0x30fef86a72ea7e1109ffeae572439995c78561ffeb968dcbd61c609efc60fdd9}{0x30fe}, \href{https://etherscan.io/tx/0xc12a4155c2c90707138e4aef8883c8f724371145823e2f661f19b93e5b3a9d6e}{0xc12a} & \href{https://etherscan.io/address/0xd3f64baa732061f8b3626ee44bab354f854877ac}{0xd3f6}  & \href{https://etherscan.io/address/0xb660cae1a59336676ea1887b15eb3c0badb90d78}{0xb660},\href{https://etherscan.io/address/0x90a7482dD7fA28865f440EC0c3B783775AC01266}{0x90a7}, \href{https://etherscan.io/address/0x2f744f784000de0b8f1a7da3f0021ad56c09ce1a}{0x2f74}  \\
\hline
42& \href{https://etherscan.io/tx/0xee02781eda4108f8fb3cca8355218961ad3a3e769b3de8a869acb8ed1654a67e}{0xee02} & \href{}{-} & \href{https://etherscan.io/address/0x94641c01a4937f2c8ef930580cf396142a2942dc}{0x9464}, \href{https://etherscan.io/address/0x5217c6923a4efc5bcf53d9a30ec4b0089f080ed0}{0x5217},\href{https://etherscan.io/address/0xe83b072433f025ef06b73e0caa3095133e7c5bd0}{0xe83b}\\
\hline
43 & \href{https://etherscan.io/tx/0xc6a5a7bc31bbc9a7530189e718f7ed96789fa65c56c3a4a08079a95074e280c8}{0xc6a5}, \href{https://etherscan.io/tx/0x22ebd267d7344780e6d63cf3a76bab57b8f8fa41cf58df1a2e1707d75d8bee89}{0x22eb}
& \href{https://etherscan.io/address/0x70cbb871e8f30fc8ce23609e9e0ea87b6b222f58}{0x70cb} ,
\href{https://etherscan.io/address/0x40aa958dd87fc8305b97f2ba922cddca374bcd7f}{0x40aa} ,
\href{https://etherscan.io/address/0x55b35bf627944396f9950dd6bddadb5218110c76}{0x55b3} 
& \href{https://etherscan.io/address/0xFacf375Af906f55453537ca31fFA99053A010239}{0xfacf} \\
\hline
44 & \href{https://etherscan.io/tx/0x0788ba222970c7c68a738b0e08fb197e669e61f9b226ceec4cab9b85abe8cceb}{0x0788}, \href{https://etherscan.io/tx/0x2aec4fdb2a09ad4269a410f2c770737626fb62c54e0fa8ac25e8582d4b690cca}{0x2aec} & \href{https://etherscan.io/address/0xb40b6c5c51ac1d16cd427f535eeca87dd4fb3f1e}{0xb40b} & \href{https://etherscan.io/address/0x5f4c21c9bb73c8b4a296cc256c0cde324db146df}{0x5f4c} \\
\hline
45 & \href{https://explorer.zksync.io/tx/0x7ac4da1ea1b0903dfabda56f713ea5e4a960a3fc34467a844d037f86ee8bfe98}{0x7ac4}, \href{https://explorer.zksync.io/tx/0x99efebacb3edaa3ac34f7ef462fd8eed85b46be281bd1329abfb215a494ab0ef}{0x99ef} & \href{https://explorer.zksync.io/address/0x7d8772DCe73cDA0332bc47451aB868Ac98F335F0}{0x7d87}, \href{https://explorer.zksync.io/address/0xC5c668DcD437b901DFE877DC99329Ac2ba338035}{0xd5c6} & \href{https://explorer.zksync.io/address/0xf1D076c9Be4533086f967e14EE6aFf204D5ECE7a}{0xf1d0} \\
\hline
46 & \href{https://bscscan.com/address/0xeade071ff23bcef312dec938ece29f7da62cf45b}{0xeade} & \href{https://bscscan.com/address/0x613cc544053812ab026d60361212cdb67b46f42f}{0x613c}  & \href{https://bscscan.com/address/0xEADe071FF23bceF312deC938eCE29f7da62CF45b}{0xeade}\\
\hline
47 & \href{https://etherscan.io/tx/0xeb87ebc0a18aca7d2a9ffcabf61aa69c9e8d3c6efade9e2303f8857717fb9eb7}{0xeb87}  & \href{https://etherscan.io/address/0x0b09c86260c12294e3b967f0d523b4b2bcdfbeab}{0x0b09}& \href{https://etherscan.io/address/0x1e8419e724d51e87f78e222d935fbbdeb631a08b}{0x1e84} \\
\hline
48 & \href{https://arbiscan.io/tx/0x44a0f5650a038ab522087c02f734b80e6c748afb207995e757ed67ca037a5eda}{0x44a0}  & \href{https://arbiscan.io/address/0x271944d9D8CA831F7c0dBCb20C4ee482376d6DE7}{0x2719}& \href{https://arbiscan.io/address/0x102be4bccc2696c35fd5f5bfe54c1dfba416a741}{0x102b} \\
\hline
49 & \href{https://arbiscan.io/tx/0x36fef881f7e9560db466a343e541072a31a07391bcd0b9bcdb6cfe8ae4616fc0}{0x36fe}& \href{https://arbiscan.io/address/0x99801433f5d7c1360ea978ea18666f7be9b3abf7#code}{0x9980} & \href{https://arbiscan.io/address/0xf2744e1fe488748e6a550677670265f664d96627}{0xf274} \\
\hline
50 & \href{https://etherscan.io/tx/0xa6f63fcb6bec8818864d96a5b1bb19e8bd85ee37b2cc916412e720988440b2aa}{0xa6f6} & \href{https://etherscan.io/address/0xb5599f568D3f3e6113B286d010d2BCa40A7745AA}{0xb559} & \href{https://etherscan.io/address/0x5061F7e6dfc1a867D945d0ec39Ea2A33f772380A}{0x5061} \\

\end{tabular}
\end{table*}

\begingroup
\renewcommand{\baselinestretch}{0.99}
\balance
\bibliographystyle{ieeetr}  
\bibliography{refs-filtered}
\endgroup



\end{document}